\documentclass[twocolumn]{revtex4-1}

\usepackage{graphicx}
\usepackage[utf8]{inputenc} 
\usepackage{bm}
\usepackage{amsmath}
\usepackage{amssymb}
\usepackage[normalem]{ulem}

\usepackage[margin=1in]{geometry}
\usepackage{xcolor}
\usepackage{dcolumn}

\begin{document}
	
	\title{Particle motion driven by non-uniform thermodynamic forces}
	
	\author{J\'er\^ome Burelbach}
	\email{de\_burel@hotmail.com}
	
	\address{Institut f\"ur Theoretische Physik, Technische Universit\"at Berlin, Hardenbergstraße 36, 10623 Berlin, Germany}
	
	\date{March 2019}
	
	\maketitle
	
	\section{Abstract}
	We present a complete reciprocal description of particle motion inside multi-component fluids that extends the conventional Onsager formulation of non-equilibrium transport to systems where the thermodynamic forces are non-uniform on the colloidal scale. Based on the dynamic length and time scale separation in suspensions, the particle flux is shown to be related to the volume-averaged coupling between the Stokes flow tensor and the thermodynamic force density acting on the fluid. The flux is then expressed in terms of thermodynamic quantities that can be computed from the interfacial properties and equation of state of the colloids. Our results correctly describe diffusion and sedimentation, and suggest that force-free phoretic motion can occur even in the absence of interfacial interactions, provided that the thermodynamic gradients are non-uniform at the colloidal surface. In particular, we derive an explicit hydrodynamic form for the phoretic force resulting from these non-uniform gradients. The form is validated by the recovery of the Henry function for electrophoresis and the Ruckenstein term for thermophoresis.
	
	\section{Introduction\label{sec:-5}}
	
	Particle motion inside multi-component systems can manifest itself in many different ways, from sedimentation and diffusion \cite{Batchelor1972,Batchelor1976,Batchelor1982,Batchelor1983} to force-free phoretic motion \cite{Ruckenstein1981,Anderson1989,Wurger2010,Parola2004} and autonomous self-propulsion.\cite{Brady2011,Bickel2014,Golestanian2005,Gaspard2018} The study of these transport phenomena relies on an understanding of how particles move when the system is not in global thermodynamic equilibrium.\cite{DeGroot1963} However, formulating a theoretical description for non-equilibrium motion is often not straightforward, as it requires a precise knowledge of the dynamic properties of all components inside the considered system.\cite{Burelbach2018b} 
	
	A rigorous theoretical description of Brownian motion was first given by Einstein,\cite{Einstein1905} while Smoluchowski pioneered the motion of charged colloids in an electric field,\cite{smoluchowski1903} a force-free interfacial phenomenon widely known as electrophoresis. Derjaguin later extended Smoluchowski's idea to gradients in temperature and chemical potentials,\cite{Derjaguin1987} thus introducing the concepts of thermophoresis and diffusiophoresis. However, these initial theories were based on assumptions that did not always apply to experimental systems. For instance, Einstein's theory of Brownian motion was restricted to ideal, non-interacting particles, whereas the theoretical description of phoretic motion was based on the so-called boundary layer approximation, where the range of interaction with the fluid is assumed to be very short compared to the particle radius. Although the latter restriction was subsequently lifted by the well-known Henry function for electrophoresis,\cite{Henry1931} such a generalisation was not immediately achieved for thermophoresis or diffusiophoresis. The similarity between all interfacial phenomena was noted by Anderson,\cite{Anderson1989} who provided a unified description of phoretic motion within the boundary layer approximation. Concerning the theory of Brownian motion, Batchelor made significant progress in describing sedimentation and diffusion of interacting particles by noting that chemical potential gradients at constant temperature are macroscopically indistinguishable from external forces.\cite{Batchelor1972,Batchelor1976,Batchelor1982,Batchelor1983} Furthermore, a connection between Batchelor's and Anderson's work was drawn by Brady, who showed that diffusiophoresis of colloids can be described from either a macroscopic or a microscopic perspective.\cite{Brady2011} 
	
	However, the motion of particles in a temperature gradient remained largely disconnected from this global picture, as Batchelor's argument based on the macroscopic equivalence between chemical forces and body forces at uniform temperature could not straightforwardly be extended to systems out of thermal equilibrium. First attempts to solve this issue were made by Dhont, who elaborated on the concept of the interfacial region and the fluid reservoir, and stressed the role of the osmotic pressure gradient as a driving force behind diffusion of interacting particles.\cite{Dhont2004a,Dhont2004} Simultaneously, Parola and Piazza managed to extend the description of thermophoresis to cases beyond the boundary layer approximation, under the assumption that the local temperature gradient around the particle remains uniform.\cite{Parola2004}  Nonetheless, the perception remained that particle motion inside a temperature gradient was to be treated differently from transport phenomena at uniform temperature, leading to the development of alternative thermodynamic approaches that seemed to disagree with the hydrodynamic treatment by Smoluchowksi and Derjaguin.\cite{Fayolle2005,Duhr2006b,Dhont2007} Pivotal work in resolving this discrepancy was done by Morthomas and Würger,\cite{Morthomas2008} who showed that the thermodynamic approach can be recovered from a hydrodynamic treatment under the assumption that the interaction range is very wide compared to the particle radius, a case also known as the H\"uckel limit.\cite{huckel1924} 
	
	Although the boundary-layer treatments by Derjaguin and Smoluchowski yielded the same result for the phoretic velocity of the particles, it was noted by Anderson that their approaches were based on entirely different arguments.\cite{Anderson1989} The root of this apparent coincidence, also known as the principle of microscopic reversibility, is the building stone of Onsager's theory of non-equilibrium thermodynamics.\cite{Onsager1931,Onsager1931a}
	Onsager's theory states that the motion of all particles in a system is determined by its rate of entropy production. The system is supposed to be at local thermodynamic equilibrium
	(LTE), consisting of small homogeneous volume elements that may individually be assumed in thermodynamic equilibrium. According to Onsager's theory, particle motion is driven by thermodynamic forces and can be described using two alternative routes, one based on a force-free argument, the other based on a reciprocal argument. The more intuitive force-free argument applied by Smoluchowski has predominantly been used for phoretic motion, as it directly determines the particle velocity from a force-free momentum balance equation. However, the involved hydrodynamic treatment is far from trivial beyond the boundary layer approximation,\cite{Parola2004,Rasuli2008} which might be a reason why many still restrict their analysis of phoretic motion to thin boundary layers.\cite{Anderson1985,Ajdari2006,Wurger2008} The key advantage of Derjaguin's reciprocal argument is that it remains straightforwardly applicable beyond the boundary layer approximation.\cite{Agar1989} More specifically, Derjaguin's argument for thermophoresis exploits the fact that the thermophoretic velocity of the particles is related to the heat flux produced when a force is applied to these particles. 
	
	Following earlier work on ionic thermophoresis by Agar $et$ $al.$,\cite{Agar1989} this reciprocal argument has recently been used by Burelbach $et$ $al.$ to formulate a unified theory for phoretic motion beyond the boundary-layer limit.\cite{Burelbach2018b} 
	In particular, this description evidenced the importance of hydrodynamic boundary conditions and showed that all phoretic phenomena can be treated equivalently. However, it assumed uniform thermodynamic forces at the particle surface, which is a prerequisite for a conventional Onsager formulation of the macroscopic heat and particle fluxes. 
	
	Here, we lift this restriction by noting that a dynamic length and time scale separation allows for a $tensorial$ (rather than scalar) coupling between different fluxes and forces inside a multi-component system. In a second step, we will make further use of this dynamic separation, in order to relate the resulting reciprocal form of the colloidal flux to the phoretic force and the osmotic pressure gradient of the colloids. An explicit hydrodynamic expression for the phoretic force is then determined using the stationary forms of the continuity equations for heat and fluid particles, and the corresponding mobilities are evaluated for electrophoresis at uniform temperature and thermophoresis in the absence of electrochemical gradients.
	
	\section{Onsager's theory of non-equilibrium thermodynamics\label{sec:-4}}
	
	We begin with a brief introduction to Onsager's theory of non-equilibrium thermodynamics. To this end, let us consider a closed multi-component system that can be partitioned into homogeneous volume elements at LTE. A volume element at a local pressure $P$ and temperature $T$ contains a large number of particles of each component, with a corresponding number density $n_i$ and chemical potential $\mu_i$. 
	
	Based on the continuity equations for heat, mass and internal energy, the rate of entropy production per volume element is given by \cite{DeGroot1963}
	\begin{eqnarray}
	\sigma_S&=&\mathbf{J}_{q}\cdot\nabla\frac{1}{T}+\sum_i\mathbf{J}_i\cdot\left(-\nabla\frac{\mu_i}{T}+\frac{1}{T}\mathbf{F}_i\right)\nonumber\\
	&& -\sum_i\mathcal{J}_iA_i -\frac{1}{T}\mathbf\Gamma:\nabla\mathbf{u}_V,\label{eq:th-9}
	\end{eqnarray}
	where $i$ is an index over all components. Here, $\mathbf{J}_{q}$ is the net heat flux through the volume element. The centre-of-mass velocity $\mathbf u_V$ of the volume element is measured relative to the system boundaries.
	The entropy produced by chemical reactions of component $i$ is represented by the term $\mathcal{J}_iA_i$, where $\mathcal{J}_i$ is the composition change and $A_i$ is the affinity of the reacting component. Viscous dissipation may occur due to a tensorial gradient in the velocity $\mathbf u_V$, which couples to the viscous stress tensor $\mathbf\Gamma$. The particle
	flux of component $i$ is defined by $\mathbf{J}_i=n_i\mathbf{v}_i$, where $\mathbf{v}_i$ is the velocity of component $i$ relative to $\mathbf{u}_V$. By definition, we have $\sum_im_i\mathbf{J}_i=0$, where $m_i$ is the corresponding particle mass.
	The force $\mathbf{F}_i$, which is directly applied to component $i$, comprises conservative body
	forces that derive from macroscopic electric ($E$) or gravitational ($g$) fields: $\mathbf F_i=\mathbf F_{E}^i+\mathbf F_g^i$.
	
	The vectorial fluxes $\mathbf{J}_{q}$ and $\mathbf{J}_i$ produce entropy by coupling to the vectorial forces $\nabla\frac{1}{T}$ and $ -\nabla\frac{\mu_i}{T}+\frac{1}{T}\mathbf{F}_i$, respectively. We shall refer to these forces as the 'thermodynamic' forces. In order to clearly distinguish the entropy produced by particle motion at uniform temperature, it is useful to transform eq. (\ref{eq:th-9}) to a basis where these thermodynamic forces are linearly independent. This can be done by writing  $\nabla\frac{\mu_i}{T}=\bar{H}_i\nabla\frac{1}{T} + \frac{1}{T}\nabla_T\mu_i$, where the notation $\nabla_T$ means that the gradient is evaluated at constant temperature. The partial molar enthalpy $\bar{H}_i$ of component $i$ is defined by $\bar{H}_i=-T^2\frac{\partial}{\partial T}\left(\frac{\mu_i}{T} \right)_{P,n_j}$. The rate of entropy production per volume element can now be expressed as
	\begin{eqnarray}
	\sigma_S&=&\mathbf{J}'_{q}\cdot\nabla\frac{1}{T}+\frac{1}{T}\sum_i\mathbf{J}_i\cdot\left( -\nabla_T\mu_i+\mathbf{F}_i\right)\nonumber\\
	& &-\sum_i\mathcal{J}_iA_i -\frac{1}{T}\mathbf\Gamma:\nabla\mathbf{u}_V,\label{eq:th-48}
	\end{eqnarray}
	where the modified heat flux $\mathbf{J}'_{q}$ is related to $\mathbf{J}_{q}$ via
	\begin{equation}
	\mathbf{J}'_{q}=\mathbf{J}_{q}-\sum_i \bar{H}_i\mathbf{J}_i.\label{eq:th-69}
	\end{equation}
	Within this chosen basis of independent thermodynamic forces, the body force $\mathbf{F}_i$ on component $i$ is indistinguishable from the chemical potential gradient at constant temperature $-\nabla_T\mu_i$ of that component. Further, eq. (\ref{eq:th-69}) shows that only the enthalpy resulting from interactions between different components contributes to the modified heat flux. 
	
	If the system is in mechanical equilibrium ($d\mathbf u_V/dt=0$) and under isotropic stress ($\mathbf\Gamma=0$), the changes in temperature and chemical potential can further be related to the change in pressure $P$ via the Gibbs-Duhem equation:\cite{DeGroot1963} $dP=sdT+\sum_in_id\mu_i$,
	where $s$ is the entropy density of the volume element. We shall refer to the pressure in the Gibbs-Duhem equation as the 'thermodynamic' pressure. In order to transform this equation into the same basis as eq. (\ref{eq:th-48}), we introduce the 'modified' entropy density $s'=s-\sum_in_i\bar{S}_i$, where $\bar{S}_i=-\left( \frac{\partial\mu_i}{\partial T}\right)_{P,n_j}$ is the partial molar entropy of component $i$.
	With the thermodynamic relations $Ts=h-\sum_{i} n_i\mu_i$ and $T\bar{S}_i=\bar{H}_i-\mu_i$, we then obtain the compensation relation between the modified entropy and enthalpy density:
	\begin{equation}
	Ts'=h'\label{eq:th-16},
	\end{equation}
	where 
	\begin{equation}
	h'=h-\sum_in_i\bar{H}_i.\label{eq:-84}
	\end{equation}
	By considering the pressure difference between two neighbouring volume elements, the Gibbs-Duhem equation can be rewritten as a thermodynamic force balance equation, given by
	
	\begin{equation}
	\nabla P=h'\frac{\nabla T}{T}+\sum_in_i\nabla_T\mu_i.\label{eq:-15}
	\end{equation}
	
	As the Curie symmetry principle forbids a coupling between fluxes and forces of different tensorial ranks inside a homogeneous volume element, Onsager's theory of non-equilibrium thermodynamics postulates linear phenomenological relations between the vectorial fluxes and thermodynamic forces, of the form \cite{Onsager1931,Onsager1931a}
	
	\begin{eqnarray}
	\mathbf{J}_{i} & = & L_{iq}\nabla\frac{1}{T}+\frac{1}{T}\sum_jL_{ij}\left( -\nabla_T\mu_j+\mathbf{F}_j\right), \label{eq:th-10}\\
	\mathbf{J}'_{q} & = & L_{qq}\nabla\frac{1}{T}+\frac{1}{T}\sum_jL_{qj}\left( -\nabla_T\mu_j+\mathbf{F}_j\right). \label{eq:th-11}
	\end{eqnarray}
	Hence, the heat and particle fluxes can couple to a temperature gradient $\nabla T$ or to any thermodynamic force $-\nabla_T\mu_j+\mathbf{F}_j$ directly acting on component $j$ at constant temperature. The scalar transport coefficients $L_{iq}$, $L_{ij}$, $L_{qq}$ and $L_{qj}$ describe the coupling of the vectorial fluxes to the thermodynamic forces and depend on the interactions between the components inside the system. Moreover, the formulation of eqs. (\ref{eq:th-10}) and (\ref{eq:th-11}) specifically relies on the assumption of LTE, which implies that internal interactions of the system cannot cause any net particle motion.\cite{Burelbach2018b} Instead, particle motion can only be induced by thermodynamic forces, which in turn derive from gradients in intensive quantities such as temperature, chemical potentials or macroscopic electric and gravitational fields.
	
	The diagonal coefficients $L_{ii}$ and $L_{qq}$ have a straightforward physical interpretation. From eq. (\ref{eq:th-11}), we can identify the thermal conductivity of the volume element as 
	\begin{equation}
	\kappa=\frac{L_{qq}}{T^2}.
	\end{equation}
	Similarly, the particle flux of component $i$ induced by a body force $\mathbf{F}_{i}$ on that component is given by
	$\mathbf{J}_{i}=n_{i}\mathbf{F}_{i}/\xi_{i}$, where $\xi_{i}$ is the friction coefficient of a particle of component $i$. As a result, the diffusive permeability $L_{ii}$ and friction coefficient $\xi_{i}$ are related via
	\begin{equation}
	L_{ii} = \frac{n_{i}T}{\xi_{i}}.
	\end{equation}
	Although the interpretation of the cross-coefficients $L_{ij}$ and $L_{iq}$ is less transparent, these coefficients must be related to the interfacial forces between different components, which give rise to force-free phoretic motion. Based on the principle of microscopic reversibility, Onsager showed that these cross-coefficients are symmetric, so that
	
	\begin{eqnarray}
	L_{ij}=L_{ji}\hspace{0.3cm}\text{and}\hspace{0.3cm}L_{qi}=L_{iq}\label{eq:-1}.
	\end{eqnarray}
	These symmetry relations, also known as the Onsager reciprocal relations, form the basis of the reciprocal approach to particle motion. More specifically, they suggest that there exists a close correspondence between the interfacial stresses responsible for phoretic motion of a component and the transport of heat and fluid induced by an applied force on that component. This is the reason why the same result for the particle flux $\mathbf J_i$ is obtained by either determining $L_{ij}$ and $L_{iq}$ from a force-free argument, or by computing $L_{ji}$ and $L_{qi}$ using the reciprocal argument.
	
	The equations presented in this section summarise the standard formulation of Onsager's theory. In arriving at eqs. (\ref{eq:th-48}), (\ref{eq:-15}), (\ref{eq:th-10}) and (\ref{eq:th-11}), it has been assumed that the thermodynamic forces are reasonably weak and uniform inside a volume element. The assumption of weak forces is generally required for the condition of LTE. However, the assumption of uniformity is a separate condition that deserves further attention. Although the system is homogeneous on the scale of a volume element, it may well be inhomogeneous on smaller scales. A volume element must be small enough to satisfy the condition of LTE, but it must also be large enough to contain a large number of particles of each component, so that all intensive thermodynamic quantities can be defined in an unambiguous manner. The minimal size of a volume element is therefore set by the typical distance between particles of the sparsest component inside the system. If the motion of all components occurs on similar length and time scales, as in molecular mixtures, it is reasonable to assume that intensive quantities only vary from one volume element to the next, such that the thermodynamic forces are uniform inside a volume element. However, this assumption must be questioned inside systems where a dynamic length and time scale separation occurs between different components.\cite{Brady2011} In this case, some intensive thermodynamic quantities remain definable well below the scale of a volume element, giving rise to local thermodynamic forces on scales where the system can no longer be considered homogeneous.
	
	\section{Motion in colloidal suspensions}
	
	The purpose of this work is to use the reciprocal argument for a general description of particle motion due to locally non-uniform thermodynamic forces. We will refer to the component of interest as the 'colloids', whereas the other components will be referred to as the 'fluid'. Let there be $N$ colloids inside a volume element of size $V$, each occupying a rigid, incompressible volume $V_c$. To simplify the notation, the Einstein convention will be used with the index $k$ for summations over the fluid components, and the index referring to the colloids will be omitted. Within this convention, the sums will always be indicated by a product containing one quantity with a lowered index and another quantity with a raised index, $e.g.$ 
	\begin{equation}
	L_{kk}a_kb^k\equiv\sum_k L_{kk}a_kb_k=L_{00}a_0b_0+L_{11}a_1b_1+...
	\end{equation}
	
\begin{figure}
		\centering{}\includegraphics[width=5.9cm,height=6.7cm]{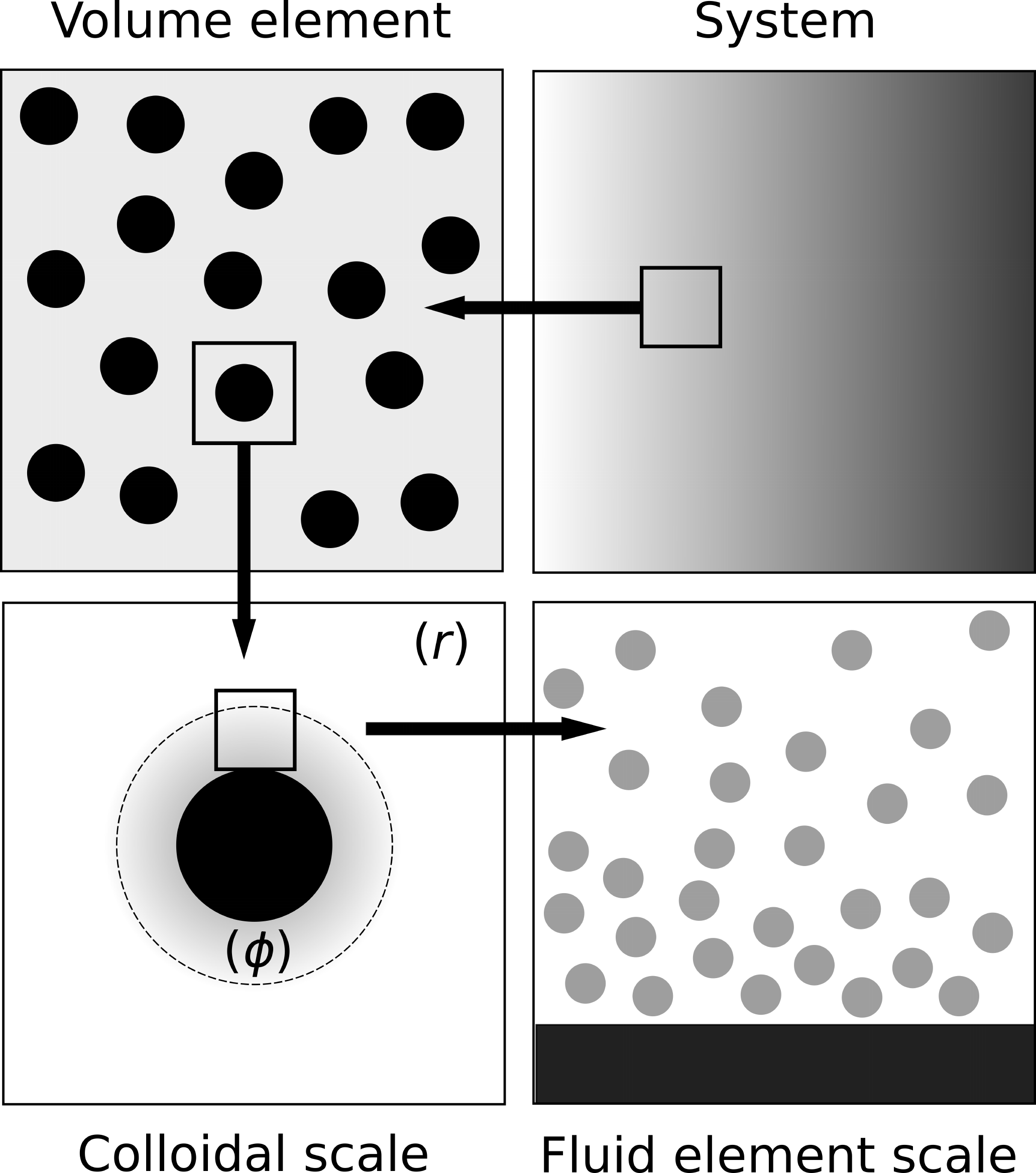}\caption{The system can be partitioned into volume elements that are at local thermodynamic equilibrium. The suspension is homogeneous on the scale of a volume element, but inhomogeneous on the colloidal scale. Due to the dynamic length and time scale separation, the fluid can be divided into an interfacial region ($\phi$) and a reservoir ($r$), and its motion can be treated hydrodynamically on the fluid element scale. The colloids are depicted by black spheres and the fluid particles by grey spheres.}\label{fig:-1}
	\end{figure}

	In order to determine the colloidal flux, we have to introduce a set of specific assumptions for the considered system. In colloidal suspensions, these assumptions derive from the dynamic length and time scale separation that occurs between the colloids and the fluid (fig. \ref{fig:-1}).\cite{Burelbach2018b} The length scale separation is based on the fact that the particles of each fluid component are very small and abundant compared to the colloids. As a result, the lower bound for $V$ is set by the mean intercolloidal volume ($cV\gtrsim 1$), so that both the colloidal concentration $c=N/V$ and colloidal flux $\mathbf J=c\mathbf v$ can be assumed uniform inside a volume element. The time scale separation occurs between the (longer) time required for the colloid to move a distance equal to its own diameter and the (shorter) time over which fluid motion reaches a steady state. This allows the use of the continuum approximation, within which fluid motion induced by the colloids can be described in terms of hydrodynamic fluid flows. The fluid thus constitutes a dynamic system made of separate fluid elements. Here, we consider a fluid that mainly consists of an electrically neutral, incompressible solvent, and that may additionally contain small (charged or uncharged) solutes. 
	
	For a description of the local dynamics, we have to distinguish between different positions inside the volume element, which might either be occupied by fluid, or by the constituent material of a colloid. A local fluid element at position $\mathbf r$ has a fluid flow velocity $\mathbf u(\mathbf r)$ (centre-of-mass velocity of the fluid relative to $\mathbf{u}_V$), but also a well-defined fluid component density $n_k(\mathbf r)$ and modified enthalpy density $h_f'(\mathbf r)$. The fluid element may further be subjected to a local temperature gradient $\nabla T(\mathbf r)$, or to local thermodynamic forces $-\nabla_T\mu_k(\mathbf r)+\mathbf F_k(\mathbf r)$ acting on its components. Although the condition of LTE requires these local thermodynamic forces to be weak, they may nonetheless be non-uniform on the colloidal scale. To first order in the gradients, we can write the thermodynamic pressure inside a fluid element as
	\begin{equation}
	\nabla P_f(\mathbf r)=h'_f(\mathbf r)\frac{\nabla T(\mathbf r)}{T}+n_k(\mathbf r)\nabla_T\mu^k(\mathbf r),\label{eq:-8}
	\end{equation}
	where $T$ is the average temperature of the volume element. The index $k$ exclusively runs over the fluid components, including the solvent and the solutes. Similarly, the body force density of a fluid element is given by
	\begin{equation}
	\mathbf f_f(\mathbf r)=n_k(\mathbf r)\mathbf F^k(\mathbf r).\label{eq:-9}
	\end{equation}
	For the following considerations, it is instructive to introduce the thermodynamic fluid force density $\bm{\mathcal F}(\mathbf r)=\mathbf f_f(\mathbf r)-\nabla P_f(\mathbf r)$, such that
	\begin{equation}
	\bm{\mathcal F}(\mathbf r)=-h'_f(\mathbf r)\frac{\nabla T(\mathbf r)}{T}+n_k(\mathbf r)\left(-\nabla_T\mu^k(\mathbf r)+\mathbf F^k(\mathbf r)\right).\label{eq:-21}
	\end{equation} 
	
	The transport properties of the fluid, such as the dynamic viscosity, electric permittivity or thermal conductivity, are set by those of the incompressible solvent and can therefore be assumed uniform inside a volume element. The solvent is also supposed to have a high dynamic viscosity, so that hydrodynamic fluid flows can be treated within the laminar Stokes regime. 
	
	A local thermodynamic gradient $\nabla Y(\mathbf r)$ usually derives from an imposed thermodynamic field $Y_b$ in the bulk ($b$) of the system, for instance by imposing different values of $Y_b$ at the system boundaries. The corresponding thermodynamic force $-\nabla Y(\mathbf r)$ on a fluid element would have a uniform value $-\nabla Y_b$ inside a volume element if the colloids were replaced by fluid. However, the fields are perturbed by the colloids if their internal transport properties differ from those of the fluid, giving rise to thermodynamic forces that are non-uniform on the colloidal scale.\cite{Epstein1929,Henry1931}
	Alternatively, a thermodynamic gradient may derive from a local source (or sink) distribution generated internally by the colloids themselves.\cite{Brady2011} In any case, the variation of the field $Y(\mathbf r)$ must be small over the volume element, as to fulfill the condition of LTE.
	
	Here, we assume that all externally imposed thermodynamic gradients $\nabla Y_b$ are applied in the same direction, and that the chemical potential of the solvent is allowed to evolve freely. This guarantees a rapid momentum relaxation inside the system, which justifies the assumption of mechanical equilibrium ($d\mathbf u_V/dt=0$) and isotropic stress ($\mathbf\Gamma=0$).\cite{Peppin2005} The net pressure of a volume element then simply coincides with the thermodynamic pressure $P$, and the momentum balance equation of a volume element reduces to
	\begin{equation}
	\mathbf f-\nabla P=0,\label{eq:-16}
	\end{equation}
	where $\mathbf f$
	is the net body force density. 
	As noted by Brady,\cite{Brady2011} this momentum balance equation is to be understood as a volume average for colloidal suspensions, meaning that it only holds if the body and pressure forces are properly summed over the volume $V$. This is particularly important when local densities and thermodynamic forces vary on the colloidal scale. As a result, the pressure gradient and body force density in eq. (\ref{eq:-16}) correspond to averages over the volume element, such that
	\begin{equation}
	\nabla P\equiv\left\langle\nabla P(\mathbf r)\right\rangle_V \ \ \text{and} \ \ \mathbf f\equiv\left\langle\mathbf f(\mathbf r)\right\rangle_V, 
	\end{equation}
	where $\nabla P(\mathbf r)$ and $\mathbf f(\mathbf r)$ are the local pressure gradient and body force density at any position inside the volume element, whether occupied by a fluid element or by the constituent material of a colloid.
	
	Before we proceed, it is useful to introduce some notation. More generally, let $\mathbf G(\mathbf r)$ be a local quantity that depends on the position inside the volume element. We write the volume average of $\mathbf G(\mathbf r)$ as 
	\begin{equation}
	\left\langle \mathbf G(\mathbf r)\right\rangle_V=\frac{1}{V}\int_N\mathbf G(\mathbf r)dV,
	\end{equation}
	where the notation $\int_N$ indicates that the integral is performed over a volume element $V$ that contains $N$ colloids. As the suspension is homogeneous on the scale of a volume element, the volume averaged gradient of $\mathbf G(\mathbf r)$ is equal to the gradient of its volume average:
	\begin{equation}
	\left\langle\nabla\mathbf G(\mathbf r)\right\rangle_V=\nabla\left\langle \mathbf G(\mathbf r)\right\rangle_V.\label{eq:-99}
	\end{equation}
	We also introduce the excluded colloidal volume function $\Theta(\mathbf r)$, defined by
	\begin{equation}
	\Theta(\mathbf r)=
	\begin{cases}
	1 & \text{inside the fluid,}\\
	0 & \text{inside the colloids}.\label{eq:-100}
	\end{cases}
	\end{equation}
	As a result, we simply have $\left\langle\Theta(\mathbf r)\right\rangle_V=1-\gamma$, where $\gamma=cV_c$ is the colloidal volume fraction. In particular, we can write
	\begin{equation}
	\left\langle\mathbf G(\mathbf r)\right\rangle_V=(1-\gamma)\left\langle\mathbf G(\mathbf r)\right\rangle_{V_f}+\gamma\left\langle\mathbf G(\mathbf r)\right\rangle_{NV_c},
	\end{equation}
	where $\left\langle\mathbf G(\mathbf r)\right\rangle_{V_f}$ represents the average of $\mathbf G(\mathbf r)$ over the volume $V_f=V-NV_c$ occupied by the fluid, and $\left\langle\mathbf G(\mathbf r)\right\rangle_{NV_c}$ is the average of $\mathbf G(\mathbf r)$ over the volume $NV_c$ occupied by the colloids. Moreover, the mean field of $\mathbf G(\mathbf r)$ inside the fluid is given by $\left\langle\mathbf G(\mathbf r)\right\rangle_{V_f}\Theta(\mathbf r)$. If $\mathbf G(\mathbf r)$ is uniform inside the fluid over the entire volume element, then $\mathbf G(\mathbf r)\Theta(\mathbf r)=\left\langle\mathbf G(\mathbf r)\right\rangle_{V_f}\Theta(\mathbf r)$. We can therefore define the perturbation of $\mathbf G(\mathbf r)$ inside the fluid as
	\begin{equation}
	\delta\mathbf G(\mathbf r)=\mathbf G(\mathbf r)\Theta(\mathbf r)-\left\langle\mathbf G(\mathbf r)\right\rangle_{V_f}\Theta(\mathbf r).\label{eq:-34}
	\end{equation}
	
	We are now ready to consider the momentum balance equation given by eq. (\ref{eq:-16}). Using eq. (\ref{eq:-15}) for the local thermodynamic pressure gradient and averaging over the volume element, the average pressure gradient can be expressed as
	\begin{eqnarray}
	\left\langle\nabla P(\mathbf r)\right\rangle_V&=&\left\langle h'(\mathbf r)\frac{\nabla T(\mathbf r)}{T}\right\rangle_V\nonumber\\
	&&+\left\langle n_k(\mathbf r)\nabla_T\mu^k(\mathbf r)\right\rangle_V+c\nabla_T\mu.\label{eq:-2}
	\end{eqnarray}
	Here, $\mu$ is the colloidal chemical potential and $h'(\mathbf r)$ is the modified enthalpy density at position $\mathbf r$. As this position might be occupied by either colloid or fluid, we can write $h'(\mathbf r)$ as
	\begin{equation}
	h'(\mathbf r)=h_f'(\mathbf r)+h_c'(\mathbf r),\label{eq:-83}
	\end{equation}
	where $h_c'(\mathbf r)$ represents the modified enthalpy density inside the colloids. Similarly, the average body force density can be written as 
	\begin{equation}
	\left\langle\mathbf f(\mathbf r)\right\rangle_V=\left\langle n_k(\mathbf r)\mathbf F^k(\mathbf r)\right\rangle_V+c\mathbf F,\label{eq:-17}
	\end{equation}
	where $\mathbf F$ is the body force on the colloids. Combining eqs. (\ref{eq:-2}) and (\ref{eq:-17}), the momentum balance equation $\left\langle\mathbf f(\mathbf r)\right\rangle_V-\left\langle\nabla P(\mathbf r)\right\rangle_V=0$ of the volume element becomes
	\begin{eqnarray}
	0&=&-\left\langle h'(\mathbf r)\frac{\nabla T(\mathbf r)}{T}\right\rangle_V\nonumber\\
	&&+\left\langle n_k(\mathbf r)\left(-\nabla_T\mu^k(\mathbf r)+\mathbf F^k(\mathbf r)\right)\right\rangle_V\nonumber\\
	&&+c\left(-\nabla_T\mu+\mathbf F\right).\label{eq:-81}
	\end{eqnarray}
	This balance equation must hold irrespective of which thermodynamic forces are applied to the system. As the temperature gradient is independent of the body forces and chemical potential gradients at constant temperature, eq. (\ref{eq:-81}) separately requires that
	\begin{equation}
	\left\langle n_k(\mathbf r)\left(-\nabla_T\mu^k(\mathbf r)+\mathbf F^k(\mathbf r)\right)\right\rangle_V+c\left(-\nabla_T\mu+\mathbf F\right)=0\label{eq:-4}
	\end{equation}
	and
	\begin{equation}
	\left\langle h'(\mathbf r)\frac{\nabla T(\mathbf r)}{T}\right\rangle_V=0.\label{eq:-82}
	\end{equation}
	These equations comply with the condition of mechanical equilibrium as derived by de Groot and Mazur for uniform gradients.\cite{DeGroot1963} Eq. (\ref{eq:-4}) is the momentum balance equation for diffusion-sedimentation systems at uniform temperature (satisfying $\left\langle\mathbf f(\mathbf r)\right\rangle_V-\left\langle\nabla_T P(\mathbf r)\right\rangle_V=0)$, whereas eq. (\ref{eq:-82}) represents the momentum balance equation for non-isothermal systems in the absence of chemical gradients or body forces (satisfying $\left\langle\frac{\partial P(\mathbf r)}{\partial T(\mathbf r)}\nabla T(\mathbf r)\right\rangle_V=0$). If the temperature gradient is uniform, then eq. (\ref{eq:-82}) reduces to $\left\langle h'(\mathbf r)\right\rangle_V=0$. In view of eq. (\ref{eq:-84}), this simply implies that the net enthalpy of a volume element is equal to the sum of the partial molar enthalpies of all particles inside it. Using eq. (\ref{eq:-83}), eq. (\ref{eq:-82}) can further be expressed as
	\begin{equation}
	\left\langle h_c'(\mathbf r)\frac{\nabla T(\mathbf r)}{T}\right\rangle_V+\left\langle h_f'(\mathbf r)\frac{\nabla T(\mathbf r)}{T}\right\rangle_V=0.\label{eq:-18}
	\end{equation} 
	
	With eqs. (\ref{eq:-4}) and (\ref{eq:-18}), we have thus related the momentum balance equation of a volume element to the thermodynamic forces acting on the components inside it, by properly averaging over the corresponding force densities. 
	
	For an Onsager formulation of the heat and particle fluxes, we now have to consider the average rate of entropy produced by the $local$ coupling of these fluxes to the corresponding thermodynamic forces. Using eq. (\ref{eq:th-48}) for the local entropy production at position $\mathbf r$ and averaging the volume element, the average rate of entropy produced by the fluxes can be expressed to first order in the gradients as
	\begin{eqnarray}
	\left\langle\sigma_S(\mathbf r)\right\rangle_V&=&-\left\langle\mathbf{J}'_{q}(\mathbf r)\cdot\frac{\nabla T(\mathbf r)}{T^2}\right\rangle_V\nonumber\\
	&&+\frac{1}{T}\left\langle\mathbf{J}_k(\mathbf r)\cdot\left( -\nabla_T\mu^k(\mathbf r)+\mathbf{F}^k(\mathbf r)\right)\right\rangle_V\nonumber\\
	&&+\frac{1}{T}\mathbf{J}\cdot\left( -\nabla_T\mu+\mathbf{F}\right),\label{eq:-65}
	\end{eqnarray}
	where $\mathbf{J}'_{q}(\mathbf r)$ is the local heat flux and $\mathbf{J}_{k}(\mathbf r)$ is the local particle flux of fluid component $k$. We recall that the colloidal flux $\mathbf J$ is uniform inside the volume element. In view of eq. (\ref{eq:-65}), the local fluxes $\mathbf{J}'_{q}(\mathbf r)$ and $\mathbf{J}_{k}(\mathbf r)$ are described by the phenomenological expressions
	\begin{eqnarray}
	\mathbf{J}'_{q}(\mathbf r) & = & -L_{qq}(\mathbf r)\frac{\nabla T(\mathbf r)}{T^2}\nonumber\\
	&&+\frac{1}{T}L_{ql}(\mathbf r)\left( -\nabla_T\mu^l(\mathbf r)+\mathbf{F}^l(\mathbf r)\right)\nonumber\\
	&&+\frac{1}{T}\mathbf L_{q}(\mathbf r)\cdot\left( -\nabla_T\mu+\mathbf F\right),\label{eq:-39}
	\end{eqnarray}
	and
	\begin{eqnarray}
	\mathbf{J}_{k}(\mathbf r) & = & -L_{kq}(\mathbf r)\frac{\nabla T(\mathbf r)}{T^2}\nonumber\\
	&&+\frac{1}{T}L_{kl}(\mathbf r)\left( -\nabla_T\mu^l(\mathbf r)+\mathbf{F}^l(\mathbf r)\right)\nonumber\\
	&&+\frac{1}{T}\mathbf L_{k}(\mathbf r)\cdot\left( -\nabla_T\mu+\mathbf F\right),\label{eq:-38}
	\end{eqnarray}
	where the indices $k$ and $l$ run over the fluid components. The last terms in eqs. (\ref{eq:-39}) and (\ref{eq:-38}), coupling to the thermodynamic force $-\nabla_T\mu+\mathbf F$ on the colloids, can be identified as the hydrodynamic contributions to the heat and fluid particle fluxes. It is important to note that the corresponding Onsager coefficients $\mathbf L_{q}(\mathbf r)$ and $\mathbf L_{k}(\mathbf r)$ have a tensorial character, since the resulting hydrodynamic fluid flows are not simply proportional to this force. In addition, we have made use of eq. (\ref{eq:-1}) by omitting the index for the colloids in the notation of $\mathbf L_{q}(\mathbf r)$ and $\mathbf L_{k}(\mathbf r)$, as Onsager's reciprocal relations imply that the order of these indices does not matter.
	
	To proceed with our reciprocal argument, we need hydrodynamic expressions for the heat and fluid particle fluxes induced by an applied force $\mathbf F$ on the colloids. Let $\mathbf u_\text{F}(\mathbf r)$ be the local fluid flow velocity induced by this force. We can then write the corresponding fluid particle flux as
	\begin{equation}
	\mathbf J_{k,\text{F}}(\mathbf r)=n_k(\mathbf r)\mathbf u_{\text{F}}(\mathbf r).\label{eq:-68}
	\end{equation}
	The force $\mathbf F$ results in a drift velocity $\mathbf v_\text{F}$ of the colloids, given by $\mathbf v_\text{F}=\mathbf F/\xi$, where $\xi$ is the friction coefficient of a colloid. Therefore, the corresponding heat flux $\mathbf J'_{q,\text{F}}(\mathbf r)$ has a contribution $h_c'(\mathbf r)\mathbf v_\text{F}$ due to the drift velocity $\mathbf v_\text{F}$ of the colloids, and a hydrodynamic contribution $h_f'(\mathbf r)\mathbf u_\text{F}(\mathbf r)$ due to the local fluid flows
	\begin{equation}
	\mathbf J'_{q,\text{F}}(\mathbf r)=h_f'(\mathbf r)\mathbf u_\text{F}(\mathbf r)+h_c'(\mathbf r)\mathbf v_\text{F}.\label{eq:-67}
	\end{equation}
	Onsager's theory requires a linear mapping between the induced fluid flow velocity $\mathbf u_\text{F}(\mathbf r)$ and the force $\mathbf F$, such that
	\begin{equation}
	\mathbf u_\text{F}(\mathbf r)=\frac{1}{\xi}\mathbf S(\mathbf r)\cdot\mathbf F,\label{eq:-20}
	\end{equation}
	where $\mathbf S(\mathbf r)$ is the symmetric Stokes flow tensor. By comparison between eqs. (\ref{eq:-39})-(\ref{eq:-38}) and (\ref{eq:-67})-(\ref{eq:-68}), we find
	\begin{equation}
	\mathbf L_k(\mathbf r) = \frac{T}{\xi}n_k(\mathbf r)\mathbf S(\mathbf r)\label{eq:-12}
	\end{equation}
	and
	\begin{equation}
	\mathbf L_q(\mathbf r) = \frac{T}{\xi}\left(h'_f(\mathbf r)\mathbf S(\mathbf r)+h'_c(\mathbf r)\mathbf 1\right),\label{eq:-41}
	\end{equation}
	where $\mathbf 1$ is the unity tensor. As the volume $V_c$ of a colloid is assumed incompressible, the incompressibility of the fluid flows requires that $\left\langle\mathbf u_{\text{F}}(\mathbf r)\right\rangle_V+\gamma\mathbf v_{\text{F}}=0$, where we recall that $\gamma=cV_c$. From this, it directly follows that
	\begin{equation}
	\left\langle\mathbf S(\mathbf r)\right\rangle_V=-\gamma\mathbf 1.\label{eq:-101}
	\end{equation}
	
	Based on Onsager's reciprocal relations, the tensors $\mathbf L_q(\mathbf r)$ and $\mathbf L_k(\mathbf r)$ also describe the coupling of the local thermodynamic forces to colloidal motion. In view of eq. (\ref{eq:-65}), the colloidal flux is therefore described by the phenomenological Onsager form
	\begin{eqnarray}
	\mathbf J &=& -\left\langle\mathbf L_{q}(\mathbf r)\cdot\frac{\nabla T(\mathbf r)}{T^2}\right\rangle_V\nonumber\\
	&&+\frac{1}{T}\left\langle\mathbf L_{k}(\mathbf r)\cdot\left( -\nabla_T\mu^{k}(\mathbf r)+\mathbf{F}^{k}(\mathbf r)\right)\right\rangle_V\nonumber\\
	&&+\frac{c}{\xi}\left( -\nabla_T\mu+\mathbf{F}\right),\label{eq:-5}
	\end{eqnarray}
	which, by using eqs. (\ref{eq:-4}), (\ref{eq:-18}), (\ref{eq:-12}) and (\ref{eq:-41}), can alternatively be written as
	\begin{equation}
	\begin{split}
	\mathbf J=\frac{1}{\xi}&\left\langle\left(\mathbf S(\mathbf r)-\mathbf 1\right)\cdot\left[-h'_f(\mathbf r)\frac{\nabla T(\mathbf r)}{T}\right.\right.\\
	&\left.\left.+n_k(\mathbf r)\left(-\nabla_T\mu^{k}(\mathbf r)+\mathbf{F}^{k}(\mathbf r)\right)\right]\right\rangle_V.\label{eq:-72}
	\end{split}
	\end{equation}
	Due to the linear response assumption of Onsager's theory, the densities $n_k(\mathbf r)$ and $h_f'(\mathbf r)$ in eq. (\ref{eq:-72}) are evaluated to zeroth order in the gradients. With eq. (\ref{eq:-21}), eq. (\ref{eq:-72}) can alternatively be expressed as
	\begin{equation}
	\mathbf J=\frac{1}{\xi}\left\langle \left(\mathbf S(\mathbf r)-\mathbf 1\right)\cdot\bm{\mathcal F}(\mathbf r)\right\rangle_V.\label{eq:-22}
	\end{equation}
	This general form of the colloidal flux constitutes the final result of our reciprocal approach, showing that the velocity of the colloids is directly related to the volume average of the product between the Stokes flow tensor and the thermodynamic fluid force density. In particular, it should be noted that the tensor $\mathbf S(\mathbf r)-\mathbf 1$ represents the Stokes flow inside the rest frame of the colloids. In the following section, we will show that eq. (\ref{eq:-22}) accounts for any colloidal transport phenomenon, be it phoretic motion or sedimentation at uniform temperature. To achieve this, we have to make further use of the dynamic length and time scale separation.
	
	\section{The interfacial region and the fluid reservoir}
	
\begin{figure}
		\centering{}\includegraphics[width=5.6cm,height=4.1cm]{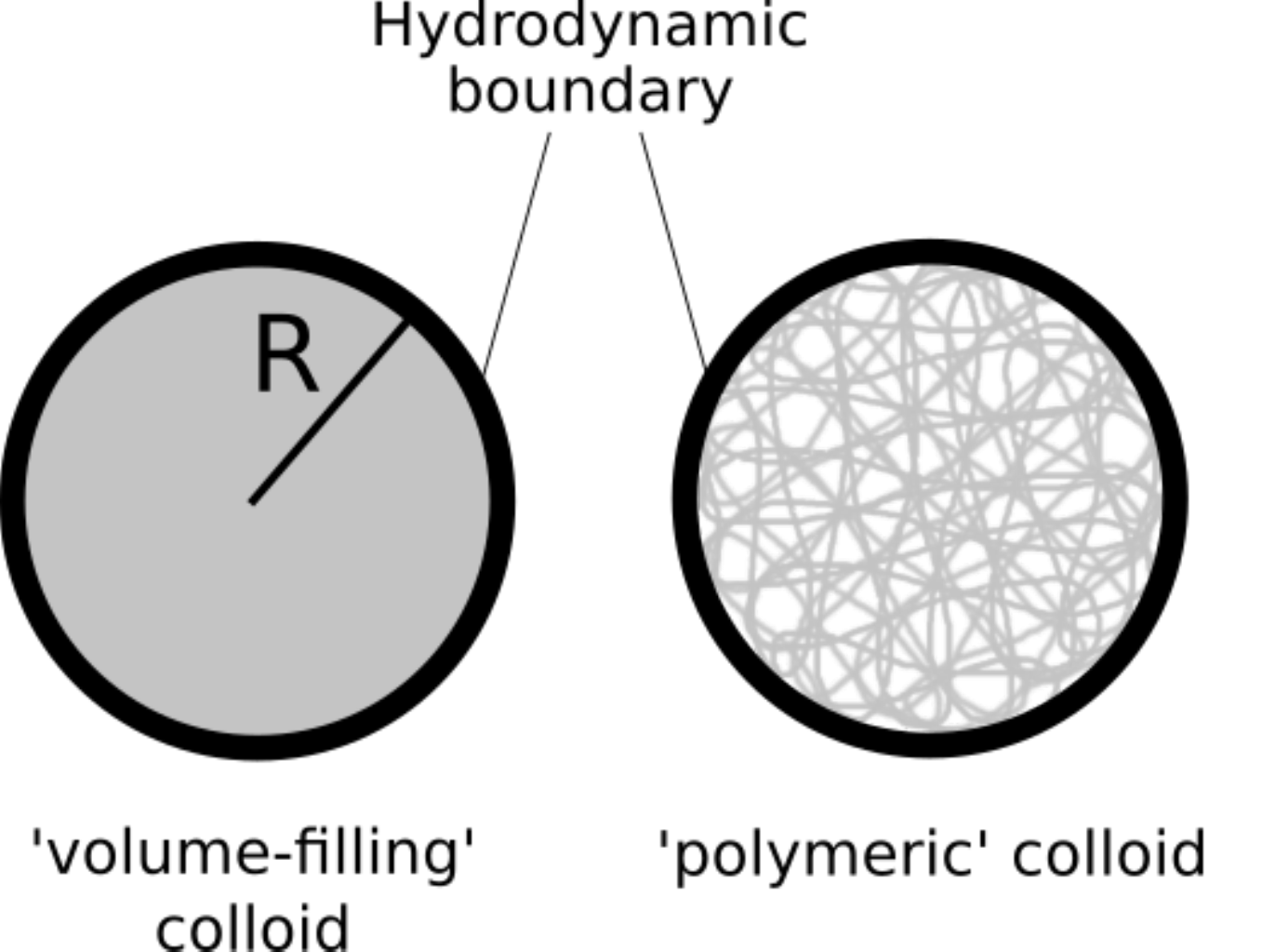}\caption{Two colloids with the same hydrodynamic boundary, but with different internal structures. The constituent material of the colloid, which occupies a volume $V_c$, is shown in grey. The volume of the left colloid coincides with the volume enclosed by its hydrodynamic boundary ($V_c=V_R$), whereas the volume of the right colloid is smaller than that of its hydrodynamic boundary ($V_c<V_R$).}\label{fig:-2}
	\end{figure}

	The dynamic length and time scale separation allows us to clearly distinguish different interactions inside the system. 
	On one hand, there are collisional interactions between the colloids and the fluid that occur on the molecular scale, such as pressure forces and viscous shear forces. Within the continuum approximation, these collisional interactions set the boundary conditions for the hydrodynamic fluid flows.\cite{Landau1987} Here, we consider colloids whose hydrodynamic boundary is approximately spherical, with a hydrodynamic radius $R$. Note that the volume $V_c$ occupied by the constituent atoms of a colloid need not be equal to, but must lie within the volume $V_R=\frac{4}{3}\pi R^3$ enclosed by its hydrodynamic boundary, so that $V_c \leqslant V_R$ (see fig. \ref{fig:-2}). The fluid particles may thus penetrate this boundary by diffusion. 
	
	On the other hand, there are specific 'interfacial' interactions between the colloids and the fluid, such as the screened electrostatic interaction known from Debye-Hückel theory,\cite{Henry1931} as well as van-der-Waals interactions or hard-core interactions. The conservative body forces deriving from these interactions lead to the build-up of a local region around the colloids that contains an interfacial excess of fluid particles and fluid enthalpy \cite{Ruckenstein1981} (fig. \ref{fig:-1}). The interfacial region ($\phi$) is in contact with a fluid reservoir ($r$) and can therefore be treated as a grand-canonical subsystem.\cite{Dhont2007} Moreover, the local densities $h'_f(\mathbf r)$ and $n_k(\mathbf r)$ of the fluid can respectively be expressed as the sum of an interfacial term and a reservoir term:
	\begin{equation}
	h'_f(\mathbf r)=h'_\phi(\mathbf r)+h_r'(\mathbf r) \ \ \text{and} \ \
	n_k(\mathbf r)=n_k^\phi(\mathbf r)+n_k^r(\mathbf r),\label{eq:-37}
	\end{equation}
	where $h'_\phi(\mathbf r)$ is the interfacial excess enthalpy density of the fluid, and $n_k^\phi(\mathbf r)$ is the interfacial excess number density of fluid component $k$. To zeroth order in the gradients, the reservoir densities $h'_r(\mathbf r)$ and $n_k^r(\mathbf r)$ are uniform, such that 
	\begin{equation}
	h'_r(\mathbf r)=h_b'\Theta(\mathbf r) \ \ \text{and} \ \
	n_k^r(\mathbf r)=n_k^b\Theta(\mathbf r),
	\end{equation}
	where $h_b'$ and $n_k^b$ are constant bulk values.
	Given that eq. (\ref{eq:-21}) is linear in the densities, the thermodynamic fluid force density splits up accordingly:
	\begin{equation}
	\bm{\mathcal F}(\mathbf r)=\bm{\mathcal F}_\phi(\mathbf r)+\bm{\mathcal F}_r(\mathbf r),\label{eq:-28}
	\end{equation}
	where
	\begin{equation}
	\bm{\mathcal F}_\phi(\mathbf r)=\mathbf f_\phi(\mathbf r)-\nabla P_\phi(\mathbf r)
	\end{equation}
	refers to the interfacial region, and
	\begin{equation}
	\bm{\mathcal F}_r(\mathbf r)=\mathbf f_r(\mathbf r)-\nabla P_r(\mathbf r)\label{eq:-70}
	\end{equation}
	to the fluid reservoir.
	
	The interfacial fluid force density $\bm{\mathcal F}_\phi(\mathbf r)$ couples the the chemical potential gradients of the solutes, but does not couple to the chemical potential gradient of the solvent, which is assumed incompressible. As the interfacial region constitutes a Grand canonical ensemble in contact with a fluid reservoir, its net free energy is given by the Grand potential $\Omega_\phi=-V\left\langle P_\phi(\mathbf r)\right\rangle_V$ and increases linearly with the number of colloids.\cite{Mori2013} Therefore, the insertion of a colloid causes the interfacial free energy of the system to change by an amount $\Omega_\phi/N$. Given that $\Omega_\phi\propto N$, this change is independent of colloidal concentration and by definition equal to the interfacial part $\mu_\phi$ of the colloidal chemical potential, so that
	$\mu_\phi=-\frac{1}{c}\left\langle P_\phi(\mathbf r)\right\rangle_V$. Taking the gradient on both sides and using eq. (\ref{eq:-99}), we obtain
	\begin{equation}
	\left\langle\nabla P_\phi(\mathbf r)\right\rangle_V=-c\nabla\mu_\phi,\label{eq:-7}
	\end{equation}
	where we note that $\nabla\mu_\phi$ is the $total$ gradient of the interfacial chemical potential.
	
	The electric force $\mathbf F_{E}^k(\mathbf r)$ on a charged fluid component stems from an electric field $\mathbf E(\mathbf r)$, such that $\mathbf F_{E}^k(\mathbf r)=e^k\Theta(\mathbf r)\mathbf E(\mathbf r)$, where $e^k$ is the corresponding fluid particle charge and $\Theta(\mathbf r)\mathbf E(\mathbf r)$ is the electric field inside the fluid. The excess charge density $\rho_e^\phi(\mathbf r)=n_k^\phi(\mathbf r)e^k$ of the interfacial region is related to a local electric potential $\varphi_E(\mathbf r)$ via $\epsilon\nabla^2\varphi_E(\mathbf r)=-\rho_e^\phi(\mathbf r)$, where $\epsilon$ is the electric permittivity of the solvent. As mentioned in section \ref{sec:-4}, the condition of LTE implies that the local electric forces deriving from $\varphi_E(\mathbf r)$ cannot induce any net colloidal motion.\cite{Burelbach2018b} Inside the fluid, the electric field $\mathbf E(\mathbf r)$ therefore exclusively derives from an electric potential $\Phi_E(\mathbf r)$ within the fluid reservoir, for instance by imposing different values of $\Phi_E$ at the system boundaries. This reservoir potential satisfies $\epsilon\nabla^2\Phi_E(\mathbf r)=-\rho_e^r(\mathbf r)$, where $\rho_e^r(\mathbf r)=n_k^r(\mathbf r)e^k$ is the local charge density of the reservoir. As the fluid reservoir is neutral everywhere to zeroth order in the gradients ($n_k^be^k=0$), the electric forces inside it cancel up to first order in the gradients:
	\begin{equation}
	n_k^b\mathbf F_{E}^k(\mathbf r)=n_k^be^k\Theta(\mathbf r)\mathbf E(\mathbf r)=0.\label{eq:-80}
	\end{equation}
	Moreover, the electric forces inside the interfacial region must balance those on the colloids, hence
	\begin{equation}
	\left\langle n_k^\phi(\mathbf r)\mathbf F_{E}^k(\mathbf r)\right\rangle_V+c\mathbf F_{E}=0\label{eq:-6}.
	\end{equation}
	
	The separation of the fluid into an interfacial region and a fluid reservoir allows us to consider the coupling between the thermodynamic forces and Stokes flows separately inside these regions. In view of eqs. (\ref{eq:-7}) and (\ref{eq:-6}), it is useful to separate out the uniform gravitational forces $\mathbf F_g^k$ from the interfacial force density $\bm{\mathcal F}_\phi(\mathbf r)$. To this end, we introduce the phoretic force density $\bm{\mathcal F}_{\text{ph}}(\mathbf r)=-\nabla P_\phi(\mathbf r)+\rho_e^\phi(\mathbf r)\mathbf E(\mathbf r)$, which is responsible for the force-free interfacial motion of the colloids. With $\rho_e^\phi(\mathbf r)\mathbf E(\mathbf r)=n_k^\phi(\mathbf r)\mathbf F_E^k(\mathbf r)$, we have
	\begin{eqnarray}
	\bm{\mathcal F}_{\text{ph}}(\mathbf r)&=&-h'_\phi(\mathbf r)\frac{\nabla T(\mathbf r)}{T}\nonumber\\
	&&+n_k^\phi(\mathbf r)\left(-\nabla_T\mu^k(\mathbf r)+\mathbf F_E^k(\mathbf r)\right),\label{eq:-30}
	\end{eqnarray}
	where $-\nabla_T\mu^k(\mathbf r)+\mathbf F_E^k(\mathbf r)$ is the 'electrochemical' force on fluid component $k$.	From this definition, it then directly follows that
	\begin{equation}
	\bm{\mathcal F}_\phi(\mathbf r)=\bm{\mathcal F}_{\text{ph}}(\mathbf r)+n_k^\phi(\mathbf r)\mathbf F_g^k.\label{eq:-29}
	\end{equation}
	With eqs. (\ref{eq:-7}) and (\ref{eq:-6}), the average phoretic force density inside the volume element can further be expressed as
	\begin{equation}
	\left\langle\bm{\mathcal F}_{\text{ph}}(\mathbf r)\right\rangle_V=-c\left(-\nabla\mu_\phi+\mathbf F_{E}\right),\label{eq:-31}
	\end{equation}
	which corresponds to a thermodynamic action-reaction law for the interfacial and electric forces acting on the colloids and the interfacial region. 
	
	To first order in the gradients, the reservoir force density is given by
	\begin{eqnarray}
	\bm{\mathcal F}_r(\mathbf r)&=&-h'_b\Theta(\mathbf r)\frac{\nabla T(\mathbf r)}{T}\nonumber\\
	&&+n_k^b\Theta(\mathbf r)\left(-\nabla_T\mu^k(\mathbf r)+\mathbf F_g^k\right),
	\end{eqnarray} 
	where we recall that $n_k^b\mathbf F_E^k(\mathbf r)=0$. For the determination of the colloidal flux, we relate the reservoir force density to the osmotic pressure of the colloids, defined by
	\begin{equation}
	\Pi=\left\langle P(\mathbf r)\right\rangle_V-\left\langle P_r(\mathbf r)\right\rangle_{V_f}.\label{eq:-25}
	\end{equation} 
	Note that this corresponds to the thermodynamic\cite{Dhont2004a} (rather than mechanical\cite{Rodenburg2017}) definition of the osmotic pressure, which allows a determination of $\Pi$ from the thermodynamic equation of state of the colloids.
	In order to relate $\bm{\mathcal F}_r(\mathbf r)$ to the osmotic pressure, we divide it into a fluid mean-field contribution and a perturbation
	\begin{equation}
	\bm{\mathcal F}_r(\mathbf r)=\left\langle \bm{\mathcal F}_r(\mathbf r)\right\rangle_{V_f}\Theta(\mathbf r)+\delta\bm{\mathcal F}_r(\mathbf r)\label{eq:-33},
	\end{equation}
	where we note that $\bm{\mathcal F}_r(\mathbf r)=\bm{\mathcal F}_r(\mathbf r)\Theta(\mathbf r)$, given that $\bm{\mathcal F}_r(\mathbf r)$ is exclusively defined inside the fluid. The fluid average $\left\langle\bm{\mathcal F}_r(\mathbf r)\right\rangle_{V_f}$ and perturbation $\delta\bm{\mathcal F}_r(\mathbf r)$ of the reservoir force density are respectively given by
	\begin{eqnarray}
	\left\langle\bm{\mathcal F}_r(\mathbf r)\right\rangle_{V_f}&=&-h'_b\frac{\left\langle\nabla T(\mathbf r)\right\rangle_{V_f}}{T}\nonumber\\
	&&+n_k^b\left(-\left\langle\nabla_T\mu^k(\mathbf r)\right\rangle_{V_f}+\mathbf F_g^k\right)\label{eq:-71}
	\end{eqnarray} 
	and
	\begin{equation}
	\delta\bm{\mathcal F}_r(\mathbf r)=-h'_b\frac{\delta\nabla T(\mathbf r)}{T}-n_k^b\delta\nabla_T\mu^k(\mathbf r),\label{eq:-53}
	\end{equation}
	where we have used $\delta\mathbf F_g^k=0$, due to the uniformity of the gravitational forces inside the fluid. Hence, a perturbation in $\bm{\mathcal F}_r(\mathbf r)$ can be caused by a perturbation in the temperature gradient, or by perturbations in the chemical potential gradients at constant temperature of the fluid components. Combining eqs. (\ref{eq:-9}), (\ref{eq:-16}), (\ref{eq:-17}) and $\mathbf f_f(\mathbf r)=\mathbf f_\phi(\mathbf r)+\mathbf f_r(\mathbf r)$, we have $\left\langle\nabla P(\mathbf r)\right\rangle_V=\left\langle\mathbf f_\phi(\mathbf r)+\mathbf f_r(\mathbf r)\right\rangle_V+c\mathbf F$. With this and $\left\langle\nabla P_r(\mathbf r)\right\rangle_{V_f}=-\left\langle \bm{\mathcal F}_r(\mathbf r)-\mathbf f_r(\mathbf r)\right\rangle_{V_f}$, taking the gradient of eq. (\ref{eq:-25}) and using eq. (\ref{eq:-99}) gives 
	\begin{equation}
	\nabla\Pi=\left\langle\mathbf f_\phi(\mathbf r)+\mathbf f_r(\mathbf r)\right\rangle_V+c\mathbf F+\left\langle \bm{\mathcal F}_r(\mathbf r)\right\rangle_{V_f}-\left\langle\mathbf f_r(\mathbf r)\right\rangle_{V_f}.\label{eq:-79}
	\end{equation}
	Using $\left\langle\mathbf f_r(\mathbf r)\right\rangle_V=\left(1-\gamma\right)\left\langle\mathbf f_r(\mathbf r)\right\rangle_{V_f}$ together with eqs. (\ref{eq:-80}) and (\ref{eq:-6}) in eq. (\ref{eq:-79}), the fluid average of the reservoir force density can thus be expressed as
	\begin{equation}
	\left\langle \bm{\mathcal F}_r(\mathbf r)\right\rangle_{V_f}=\nabla\Pi-c\mathbf F_g+\gamma n_k^b\mathbf F_g^k-\left\langle n_k^\phi(\mathbf r)\right\rangle_V\mathbf F_g^k.\label{eq:-32}
	\end{equation}
	Moreover, a comparison between eqs. (\ref{eq:-71}) and (\ref{eq:-32}) yields 
	\begin{eqnarray}
	\nabla_T\Pi-c\mathbf F_g&=&n_k^b\left(-\left\langle\nabla_T\mu^k(\mathbf r)\right\rangle_{V_f}+\left(1-\gamma\right)\mathbf F_g^k\right)\nonumber\\
	&&+\left\langle n_k^\phi(\mathbf r)\right\rangle_V\mathbf F_g^k
	\end{eqnarray}
	and
	\begin{equation}
	h_b'=\frac{T\partial\Pi}{\partial\left\langle T(\mathbf r)\right\rangle_{V_f}}.\label{eq:-52}
	\end{equation}
	
	Having related $\bm{\mathcal F}_\phi(\mathbf r)$ to the phoretic force density and $\bm{\mathcal F}_r(\mathbf r)$ to the osmotic pressure gradient, the thermodynamic force density $\bm{\mathcal F}(\mathbf r)=\bm{\mathcal F}_\phi(\mathbf r)+\bm{\mathcal F}_r(\mathbf r)$ can now be re-expressed in terms of these quantities. Substituting eqs. (\ref{eq:-29}), (\ref{eq:-33}) and (\ref{eq:-32}) into eq. (\ref{eq:-28}), we obtain
	\begin{equation}
	\begin{split}
	\bm{\mathcal F}(\mathbf r)&=\\
	&\bm{\mathcal F}_{\text{ph}}(\mathbf r)+n_k^\phi(\mathbf r)\mathbf F_g^k\nonumber\\
	&+\Theta(\mathbf r)\left[\nabla\Pi-c\mathbf F_g+\gamma n_k^b\mathbf F_g^k-\left\langle n_k^\phi(\mathbf r)\right\rangle_V\mathbf F_g^k\right]\nonumber\\
	&+\delta\bm{\mathcal F}_r(\mathbf r).\label{eq:-35}
	\end{split}
	\end{equation}
	Using this result for $\bm{\mathcal F}(\mathbf r)$ in eq. (\ref{eq:-22}), and noting that eqs. (\ref{eq:-100}) and (\ref{eq:-101}) give $\left\langle\Theta(\mathbf r)\left(\mathbf S(\mathbf r)-\mathbf 1\right)\right\rangle_V=-\mathbf 1$, the colloidal flux finally takes the form
	\begin{eqnarray}
	\mathbf J&=&\frac{1}{\xi}\left\langle \left(\mathbf S(\mathbf r)-\mathbf 1\right)\cdot\left( \bm{\mathcal F}_{\text{ph}}(\mathbf r)+\delta\bm{\mathcal F}_r(\mathbf r)\right) \right\rangle_V\nonumber\\
	&&+\frac{1}{\xi}\left(-\nabla\Pi+c\mathbf F_g\right)-\frac{c}{\xi}V_cn_k^b\mathbf F_g^k\nonumber\\
	&&+\frac{1}{\xi}\left\langle n_k^\phi(\mathbf r)\mathbf S(\mathbf r)\right\rangle_V\cdot\mathbf F_g^k.\label{eq:-36}
	\end{eqnarray}
	This expression of the colloidal flux is particularly useful for theoretical computations as it only refers to a hydrodynamic coupling inside the regions containing an interfacial excess, or those where the thermodynamic forces are non-uniform. Furthermore, the osmotic pressure gradient $\nabla\Pi$ can be computed if the colloidal equation of state is known.
	The term $-V_cn_k^b\mathbf F_g^k$ in eq. (\ref{eq:-36}) represents the buoyant force, which scales with the number of fluid particles $V_cn_k^b$ 'displaced' by the volume $V_c$ of a colloid. The last term in eq. (\ref{eq:-36}) is a hydrodynamic contribution stemming from the gravitational forces on the solutes inside the interfacial region, which could in principle be determined once the Stokes tensor and the interfacial excess densities are known. However, as the solutes are usually much lighter and sparser than the solvent, it is reasonable to neglect the weight of the solutes, so that $\left\langle n_k^\phi(\mathbf r)\mathbf S(\mathbf r)\right\rangle_V\cdot\mathbf F_g^k\approx \mathbf 0$ and $-\gamma n_k^b\mathbf F_g^k\approx-\gamma n_0^b\mathbf F_{g,0}$, where $n_0^b\mathbf F_{g,0}$ is the gravitational force density on the solvent.
	The first term in eq. (\ref{eq:-36}) represents the force-free phoretic motion of the colloids, which we denote by
	\begin{equation}
	\mathbf J_{\text{ph}}=\frac{1}{\xi}\left\langle \left(\mathbf S(\mathbf r)-\mathbf 1\right)\cdot\left( \bm{\mathcal F}_{\text{ph}}(\mathbf r)+\delta\bm{\mathcal F}_r(\mathbf r)\right) \right\rangle_V.
	\end{equation}
	
	To test the validity of eq. (\ref{eq:-36}), it is instructive to look at two well-known limiting cases. First, let us consider the case of uncharged colloids at uniform temperature, inside a fluid where all components are incompressible and all thermodynamic forces are uniform. This implies that $\bm{\mathcal F}_{\text{ph}}(\mathbf r)=\delta\bm{\mathcal F}_r(\mathbf r)=\mathbf 0$. From eq. (\ref{eq:-36}), we then recover the standard form of the diffusion-sedimentation flux \cite{Peppin2005}
	\begin{equation}
	\mathbf J=\frac{1}{\xi}\left(-\nabla_T\Pi+c\mathbf F_g\right)-\frac{c}{\xi}V_cn_k^b\mathbf F_g^k.
	\end{equation}
	In section \ref{sec:-5}, we briefly mentioned the thermodynamic approach to thermophoresis.\cite{Fayolle2005,Duhr2006b,Dhont2007} More generally, a thermodynamic approach is only concerned with the interfacial contribution to phoretic motion and neglects the hydrodynamic coupling to the interfacial force density, which amounts to setting $\delta\bm{\mathcal F}_r(\mathbf r)=\mathbf 0$ and $\mathbf S(\mathbf r)=\mathbf 0$ in eq. (\ref{eq:-36}).  In view of eq. (\ref{eq:-31}), eq. (\ref{eq:-36}) then reduces to
	\begin{equation}
	\mathbf J=-\frac{c}{\xi}\nabla\mu_\phi+\frac{c}{\xi}\mathbf F_{E}+\frac{1}{\xi}\left(-\nabla\Pi+c\mathbf F_g\right)-\frac{c}{\xi}V_cn_k^b\mathbf F_g^k,
	\end{equation}
	which corresponds to the thermodynamic form of the colloidal flux. However, neglecting the hydrodynamic coupling is generally not a valid assumption for colloids, since the interfacial excess is often primarily located within a thin outer layer around the hydrodynamic boundary. The opposite case, where the hydrodynamic boundary may be treated as point-like compared to the outer interfacial region, rarely occurs in experimental systems and would only hold for small nanoparticles with a very wide interaction range.\cite{Burelbach2018b} 
	
	\section{Phoretic motion in the dilute limit\label{sec:-6}}
	
	After a successful comparison to limiting results that ignore the local hydrodynamic coupling in eq. (\ref{eq:-36}), we will now focus on the force-free phoretic contribution to colloidal motion. To this end, we introduce the phoretic force $\mathbf F_{\text{ph}}=\xi\mathbf J_{\text{ph}}/c$, such that
	\begin{equation}
	\mathbf F_{\text{ph}}=\frac{1}{c}\left\langle \left(\mathbf S(\mathbf r)-\mathbf 1\right)\cdot\left(\bm{\mathcal F}_{\text{ph}}(\mathbf r)+\delta\bm{\mathcal F}_r(\mathbf r)\right) \right\rangle_V.\label{eq:-62}
	\end{equation}
	First of all, we note that phoretic motion comprises two different contributions, one coupling to the interfacial region ($\bm{\mathcal F}_{\text{ph}}(\mathbf r)$), and another one related to the perturbation of the reservoir force density ($\delta\bm{\mathcal F}_r(\mathbf r))$. The first contribution corresponds to the conventional perception of phoretic motion as an interfacial phenomenon.\cite{Ruckenstein1981,Derjaguin1987,Anderson1989} However, little is known about the second contribution, which suggests that phoretic motion can occur even in the absence of interfacial excess, under the condition that the local thermodynamic forces are non-uniform. 
	
	To understand this, it is sufficient to consider a single colloid inside an infinitely large fluid at uniform temperature, made of an incompressible solvent ($k=0$) and containing a neutral solute component ($k=1$). The system is subjected to a solute chemical potential gradient $\nabla_T\mu_1^b$. For the sake of simplicity, we assume that gravity is absent. If the chemical potential of the solvent remains unperturbed by the colloid, the local fluid pressure gradient at its surface is given by $\nabla_T P_f(\mathbf r)=n_1(\mathbf r)\nabla_T\mu_1(\mathbf r)+n_0^b\nabla_T\mu_0^b$. As the solvent allows for a net pressure equilibration in the bulk ($b$) of the system, its chemical potential gradient is fixed by the condition $ \nabla P_f^b=n_1^b\nabla_T\mu_1^b+n_0^b\nabla_T\mu_0^b=0$. The local fluid pressure gradient can therefore be expressed as $\nabla_T P_f(\mathbf r)=n_1(\mathbf r)\nabla_T\mu_1(\mathbf r)-n_1^b\nabla_T\mu_1^b$. In the conventional treatment of diffusiophoresis as an interfacial phenomenon, the solute chemical potential gradient would be assumed uniform everywhere around the colloid ($\nabla_T\mu_1(\mathbf r)=\nabla_T\mu_1^b$), yielding $\nabla_T P_f(\mathbf r)=n_1^\phi(\mathbf r)\nabla_T\mu_1^b$, where $n_1^\phi(\mathbf r)=n_1(\mathbf r)-n_1^b$ is the interfacial excess density of the solute. However, a fluid pressure gradient also arises in the absence of interfacial excess ($n_1(\mathbf r)=n_1^b$) if the colloid modifies the solute chemical potential gradient ($\nabla_T\mu_1(\mathbf r)\neq\nabla_T\mu_1^b$), giving $\nabla_T P_f(\mathbf r)=n_1^b\left(\nabla_T\mu_1(\mathbf r)-\nabla_T\mu_1^b\right)=n_1^b\delta\nabla_T\mu_1(\mathbf r)$. In both cases, this fluid pressure gradient can lead to diffusiophoretic motion.
	
	In order to obtain explicit expressions for the Stokes flow tensor and the local thermodynamic force fields, we will consider the limit of high dilution $\gamma\leqslant cV_R\ll 1$, assuming that interfacial excess densities and local force perturbations decay rapidly over a typical distance $c^{-1/3}$ between colloids. 
	As inter-colloidal interactions can be neglected in the dilute limit, the osmotic pressure of the colloids is given by the ideal expression $\Pi=ck_B\left\langle T(\mathbf r)\right\rangle_{V_f}$,\cite{Einstein1905} such that
	\begin{equation}
	\nabla\Pi=k_BT\nabla c+ck_B\left\langle\nabla T(\mathbf r)\right\rangle_{V_f}.\label{eq:-59}
	\end{equation}
	Comparison between eqs. (\ref{eq:-59}) and (\ref{eq:-52}) further yields
	\begin{equation}
	h'_b=-ck_BT.\label{eq:-51}
	\end{equation}
	The modified bulk enthalpy density $h'_b$ of the fluid is thus expected to vanish when the colloidal concentration goes to zero, and can therefore be identified as a collective effect. Unlike the aforementioned example of diffusiophoresis, a local perturbation of the temperature gradient alone will therefore not contribute to thermophoretic motion in the single-colloid limit ($c=0$).
	
	In the dilute limit, each colloid undergoes phoretic motion on its own, uninfluenced by the other colloids. As a result, the interfacial excess densities and force perturbations in the direct vicinity of one colloid only depend on the position relative to that colloid. It is hence sufficient to consider a single colloid in the middle of a volume element, with the reference point for $\mathbf r$ conveniently chosen at the centre of its hydrodynamic boundary.
	The expression for the phoretic force then reduces to
	\begin{equation}
	\mathbf F_{\text{ph}}=\int_1\left(\mathbf S(\mathbf r)-\mathbf 1\right)\cdot\left(\bm{\mathcal F}_{\text{ph}}(\mathbf r)+\delta\bm{\mathcal F}_r(\mathbf r)\right)dV,\label{eq:-73}
	\end{equation}
	where the notation $\int_1$ indicates that the integral is performed over a volume element containing a single colloid. Let us denote the temperature gradient and electrochemical gradients occurring in eq. (\ref{eq:-30}) by
	\begin{equation}
	\nabla Y(\mathbf r)\equiv\left\{k_B\nabla T(\mathbf r),\nabla_T\mu_k(\mathbf r)-e_k\mathbf E(\mathbf r)\right\},\label{eq:-50}
	\end{equation}
	and accordingly, the densities by 
	\begin{equation}
	\rho(\mathbf r)\equiv\left\{\frac{h'_f(\mathbf r)}{k_BT},n_k(\mathbf r)\right\},
	\end{equation}
	where $k_B$ is the Boltzmann constant. Using the same notation for interfacial and reservoir densities as in the previous section, we can write  
	\begin{equation}
	\rho(\mathbf r)=\rho_\phi(\mathbf r)+\rho_r(\mathbf r),
	\end{equation}
	with $\rho_r(\mathbf r)=\rho_b\Theta(\mathbf r)$ to zeroth order in the gradients. In view of eq. (\ref{eq:-73}), the phoretic force $\mathbf F_Y$ resulting from the thermodynamic gradient $\nabla Y(\mathbf r)$ can now be expressed as
	\begin{equation}
	\mathbf F_Y=-\int_1\left(\mathbf S(\mathbf r)-\mathbf 1\right)\cdot\left(\rho_\phi(\mathbf r)\nabla Y(\mathbf r)+\rho_b\delta\nabla Y(\mathbf r)\right)dV.\label{eq:-26}
	\end{equation}
	
	For a colloid with a spherical hydrodynamic boundary, the Stokes flow tensor has a well-known analytical solution, given by \cite{Landau1987,Barber2000}
	\begin{equation}
	\mathbf S(\mathbf r)=\frac{3}{4}\frac{R}{r}\left[a\left(\mathbf 1+\hat {\mathbf r}\hat {\mathbf r}\right)-\left(a-\frac{2}{3}\right)\frac{R^2}{r^2}\left(3\hat {\mathbf r}\hat {\mathbf r}-\mathbf 1\right)\right]\label{eq:-42}
	\end{equation}
	if $r\geqslant R$, and $\mathbf S(\mathbf r)=\mathbf 1$ if $r<R$. Here, $r=|\mathbf r|$ and $\hat {\mathbf r}=\mathbf r/|\mathbf r|$. The slip parameter $a$ takes the value $1$ for a non-slip boundary condition and the value $2/3$ for a perfect-slip boundary condition. Moreover, the friction coefficient of a colloid is given by Stokes' law
	\begin{equation}
	\xi=6\pi a\eta R,\label{eq:-57}
	\end{equation}
	where $\eta$ is the dynamic viscosity of the solvent. As $\mathbf S(\mathbf r)-\mathbf 1=\mathbf 0$ within the hydrodynamic boundary, only the region outside the boundary contributes to the integral in eq. (\ref{eq:-73}). The outer interfacial region surrounding the hydrodynamic boundary is also referred to as the interfacial layer. The effective width $\lambda$ of the interfacial layer is set by the steepness of the specific interaction potential outside the sphere. The layer is termed 'thin' if the potential decays rapidly over a distance small compared to the hydrodynamic radius of the colloid ($R\gg \lambda$), and 'wide' otherwise ($R\ll \lambda$).
	
	The thermodynamic fields $Y(\mathbf r)$ are determined by the continuity equations for heat and fluid particles. Due to the dynamic length and time scale separation, these equations can be used in their stationary forms, which are respectively given by $\nabla\cdot\mathbf J'_q(\mathbf r)=\sigma_q(\mathbf r)$ and $\nabla\cdot\mathbf J_k(\mathbf r)=\sigma_k(\mathbf r)$. The heat and fluid source densities $\sigma_q(\mathbf r)$ and $\sigma_k(\mathbf r)$ generated by the colloids can give rise to phoretic motion in the absence of any externally applied gradients. In this case, the colloids are said to be 'active', and the resulting motion is also known as self-phoretic motion.\cite{Gaspard2018} We will therefore refer to the colloids as 'passive' if their phoretic motion is exclusively driven by externally applied thermodynamic gradients.
	
	\subsection{The hydrodynamic form of the phoretic force for passive colloids}
	
	We now investigate the hydrodynamic form of eq. (\ref{eq:-26}) for passive colloids. In what follows, we model the inner region of the hydrodynamic boundary of a colloid as a homogeneous medium with uniform transport properties, which may differ from those of the fluid. To determine the thermodynamic fields, we follow Anderson \cite{Anderson1989} by reducing the continuity equations for heat and fluid particles to a set of Laplace equations, of the form
	\begin{equation}
	\nabla\cdot y(\mathbf r)\nabla Y(\mathbf r)=0, \ \ y(\mathbf r)=
	\begin{cases}
	y_\text{out} & \text{if}\ r\geqslant R \\
	y_\text{in} & \text{if}\ r<R,
	\end{cases}\label{eq:-40}
	\end{equation}
	where
	\begin{equation}
	y(\mathbf r)\equiv\left\{\kappa(\mathbf r),\frac{L_{kk}(\mathbf r)}{T}\right\}.
	\end{equation}
	Hence, the relevant transport coefficients for the temperature field and electrochemical fields are the thermal conductivity $\kappa(\mathbf r)$ and the diffusive permeabilities $L_{kk}(\mathbf r)/T$, respectively. Knowing that $\nabla Y(\mathbf r)$ tends to the applied bulk gradient $\nabla Y_b$ far away from the colloid, the Laplace equation can be solved for a sphere of radius $R$ and transport coefficient $y_{\text{in}}$ embedded in a fluid medium with a transport coefficient $y_{\text{out}}$. From the solution, which is given in appendix \ref{sec:-2}, the tensor $\mathbf C_{y'}(\mathbf r)$, such that 
	\begin{equation}
	\nabla Y(\mathbf r)=\mathbf C_{y'}(\mathbf r)\cdot\nabla Y_b,\label{eq:-103}
	\end{equation}
	can be identified as
	\begin{equation}
	\mathbf C_{y'}(\mathbf r)=
	\begin{cases}
	\mathbf 1+y'\frac{R^3}{r^3}\left(3\hat {\mathbf r}\hat {\mathbf r}-\mathbf 1\right), &r\geqslant R \\
	\\
	\left(1-y'\right)\mathbf 1, &r<R.\label{eq:-97}
	\end{cases}
	\end{equation}
	The constant $y'$ is given by the ratio
	\begin{equation}
	y'=\frac{y_\text{in}-y_\text{out}}{y_\text{in}+2y_\text{out}}.
	\end{equation}
	Here, we will refer to this ratio as the 'Clausius-Mossotti' factor, due to its resemblance with the ratio of electric permittivities bearing the same name. 
	
	For our model calculations, we consider a colloid with a spherically symmetric surface charge distribution, which implies that the interfacial excess densities $\rho_\phi(r)$ only depend on the radial distance $r$ from its centre, to zeroth order in the gradients. Based on the rotational symmetry around the direction of the bulk gradient $\nabla Y_b$, the following expression can then be derived for the contribution $\mathbf F_{Y}$ to the phoretic force (see appendix \ref{sec:-2}):
	\begin{equation}
	\mathbf F_{Y} =-\nabla Y_b\int_R^\infty\rho_\phi(r) B_{y'}(r) 4\pi r^2dr-y'V_R\rho_b\nabla Y_b,\label{eq:-46}
	\end{equation}
	where the dimensionless function $B_{y'}(r)$ is given by
	\begin{equation}
	B_{y'}(r)=-1+a\frac{R}{r}+y'\left[\frac{a}{2}\frac{R^4}{r^4}+\left(1-\frac{3a}{2}\right)\frac{R^6}{r^6}\right].\label{eq:-55}
	\end{equation}
	
	The first term in eq. (\ref{eq:-46}) involves an integral over the interfacial layer. For uniform thermodynamic forces ($y'=0$), this contribution reduces to the initial reciprocal approach by Burelbach $et$ $al.$ ($B_{y'}(r)=-1+aR/r$),\cite{Burelbach2018b} which has recently been applied to charged systems \cite{Burelbach2019} and validated by means of computer simulations.\cite{Burelbach2018} If the interaction range $\lambda$ is very small compared to the colloidal radius $R$, we have $\varepsilon=r-R\ll R$, and a leading order expansion in the small parameter $\varepsilon/R$ yields
	\begin{equation}
	r^2B_{y'}(r)\approx-R^2(1-a)(1-y')-(2-a+4y'-5ay')R\varepsilon.\label{eq:-49}
	\end{equation}
	For a stick boundary condition ($a=1$), this reduces to $r^2B_{y'}(r)=-(1-y')R\varepsilon$, and the contribution from the interfacial layer takes the limiting form
	\begin{equation}
	\mathbf F_{Y,\text{layer}}\approx\nabla Y_b(1-y')4\pi R\int_0^\infty \varepsilon\rho_\phi(\varepsilon)d\varepsilon.
	\end{equation}
	Noting that $1-y'=3y_{\text{out}}/(y_{\text{in}}+2y_{\text{out}})$, this exactly coincides with the well-known Smoluchowski-Derjaguin expression of the phoretic force,\cite{smoluchowski1903,Derjaguin1987} which is the widely accepted result for phoretic motion within the boundary layer approximation.\cite{Anderson1989,Wurger2010,Brady2011} However, if there is partial slip at the boundary ($\frac{2}{3}\leqslant a<1$), the first term in eq. (\ref{eq:-49}) is dominant and we get
	\begin{equation}
	\mathbf F_{Y,\text{layer}}\approx\nabla Y_b(1-a)(1-y')4\pi R^2\int_0^\infty\rho_\phi(\varepsilon)d\varepsilon,
	\end{equation}
	where $4\pi R^2\int_0^\infty\rho_\phi(\varepsilon)d\varepsilon$ is the net interfacial excess inside the layer. As the interfacial excess scales with $\sim R^2\lambda$, this result is by an order $R/\lambda$ larger than the Smoluchowski-Derjaguin expression for a stick boundary, which explains why the phoretic velocity rapidly increases with hydrodynamic slip for colloids with thin interfacial layers.\cite{Ajdari2006}
	
	The second term in eq. (\ref{eq:-46}), which stems from the force perturbation, scales with the Clausius-Mossotti factor $y'$ and is (rather unexpectedly) independent of the hydrodynamic boundary condition. Due to its proportionality to $V_R\rho_b$, it has resemblance with the buoyant force, which instead scales with $V_c\rho_b$. To our knowledge, this term has not been discussed in literature so far. We therefore conclude that a local perturbation in the temperature or fluid chemical potentials alone may be sufficient to give rise to a phoretic force. 
	
	To illustrate this, we reconsider our previous example of diffusiophoretic motion of a single colloid inside an infinitely large fluid, due to an applied chemical potential gradient $\nabla_T\mu_1^b$ of a neutral solute component ($k=1$). We again assume that gravity is absent, but now also allow for a perturbation of the solvent chemical potential ($k=0$). Applying eq. (\ref{eq:-46}) in the absence of interfacial solute excess ($n_1^\phi(\mathbf r)=0$), the diffusiophoretic force just reduces to
	\begin{eqnarray}
	\mathbf F_{\text{ph}}&=&\mathbf F_{\mu_1}+\mathbf F_{\mu_0}\\\nonumber
	&=&-L'_{11} V_Rn_1^b\nabla_T\mu_1^b-L'_{00}V_Rn_0^b\nabla_T\mu_0^b.
	\end{eqnarray}
	Note that the factor $1/T$ in the diffusive permeability $L_{kk}/T$ drops out of the corresponding Clausius-Mossotti factor, which can therefore just be written as $L'_{kk}=(L_{kk,\text{in}}-L_{kk,\text{out}})/(L_{kk,\text{in}}+2L_{kk,\text{out}})$.
	Using the condition of uniform bulk pressure $n_1^b\nabla_T\mu_1^b+n_0^b\nabla_T\mu_0^b=0$ to eliminate the solvent chemical potential, we obtain
	\begin{equation}
	\mathbf F_{\text{ph}}=-\left(L'_{11}-L'_{00}\right)V_Rn_1^b\nabla_T\mu_1^b.
	\end{equation}
	This result shows that diffusiophoresis in the absence of interfacial interactions relies on a notable difference $L'_{11}-L'_{00}$ between the relative boundary permeabilities of the solute and the solvent. Let us briefly consider a colloid whose boundary is 'transparent' to solvent diffusion ($L_{00,\text{in}}=L_{00,\text{out}}$), in which case the solvent chemical potential is not perturbed ($L'_{00}=0$). If the solute is treated within the Poisson-Boltzmann-Debye-Hückel approximation, we further have $n_1^b\nabla_T\mu_1^b=k_BT\nabla n_1^b$ and $L_{11}=n_1^bT/\xi_1\propto D_1$, where $D_1$ is the diffusion coefficient of the solute. The corresponding diffusiophoretic mobility $M_{\mu_1}$ of the colloid, defined by $\mathbf F_{\text{ph}}=-\xi M_{\mu_1}\nabla n_1^b$, can then be identified as
	\begin{equation}
	M_{\mu_1}=\frac{1}{\xi}\frac{D_{1,\text{in}}-D_{1,\text{out}}}{D_{1,\text{in}}+2D_{1,\text{out}}}V_Rk_BT.\label{eq:-104}
	\end{equation}
	For a colloid whose hydrodynamic boundary is impermeable to the solute ($D_{1,\text{in}}=0$), we obtain $M_{\mu_1}=-V_Rk_BT/(2\xi)$, meaning that the colloid will be subjected to a diffusiophoretic force that tends to pull it up the applied solute chemical potential gradient. It should however be noted that the contribution given by eq. (\ref{eq:-104}) is unlikely to be observed for conventional suspensions, as most colloids are either impermeable or equally permeable to diffusion of the solvent and the solutes. Although the design of such a diffusively semi-permeable colloid may well be within the reach of current fabrication techniques, eq. (\ref{eq:-104}) could instead be verified by means of MPCD simulations,\cite{Malevanets1999} which allow for a fine tuning of inner and outer transport properties.\cite{Burelbach2018} 
	
	In arriving at eq. (\ref{eq:-46}), we have assumed that the continuity equations for heat and fluid particles can be reduced to a set of Laplace equations, but this assumption has not been justified so far. As shown in appendix \ref{sec:-1}, this reduction requires that the solutes be treated within the Poisson-Boltzmann-Debye-Hückel approximation, which will be introduced more rigorously in the following section. In particular, the set of Laplace equations defined by eqs. (\ref{eq:-50}) and (\ref{eq:-40}) are only recovered if temperature gradients and electrochemical gradients do not occur simultaneously (see appendix \ref{sec:-1}). In the following model calculations for passive colloids, we will therefore consider two specific cases: Electrophoresis at uniform temperature, and thermophoresis in the absence of electrochemical gradients.
	
	\section{Model calculation for charged colloids: Electro- and thermophoresis\label{sec:-7}}
	
	To further validate eq. (\ref{eq:-46}), we apply our result for the phoretic force in the dilute limit to the motion of charged colloids suspended in an aqueous electrolyte solution. Phoretic motion of charged colloids has attracted a lot of attention in recent years,\cite{Morthomas2008,Fayolle2008,Wurger2008,Khair2009} and has been described beyond the boundary-layer limit from both a force-free \cite{Dhont2008,Rasuli2008} and reciprocal \cite{Burelbach2019} perspective. However, these approaches all assumed uniform gradients at the colloidal surface.\cite{Dhont2008,Rasuli2008,Burelbach2019} It is therefore instructive to evaluate eq. (\ref{eq:-46}) for charged colloids, as the resulting expressions should apply beyond the boundary layer approximation even when the thermodynamic gradients are non-uniform.
	
	The evaluation of eq. (\ref{eq:-46}) obviously requires knowledge of the densities $\rho_\phi(r)$ and $\rho_b$. However, deriving explicit expressions for the interfacial excess densities $n_k^\phi(\mathbf r)$ and $h_\phi'(\mathbf r)$ remains one of the main challenges in the theoretical description of phoretic motion inside non-ideal fluids. Although simple theoretical models, such as the Born theory for polar solvents, have previously been used to account for the solvation enthalpy of a colloid,\cite{Fayolle2008,Rasuli2008,Burelbach2019} it is currently still unclear to what extent these models apply to solid-liquid interfaces where the solvent molecules may no longer be freely polarisable. In order to circumvent this issue, interfacial colloid-solvent interactions will be ignored here, whereas the charged solutes (ions and counterions) will be described using the Poisson-Boltzmann-Debye-Hückel approximation. The solutes are thus treated as point-like particles that only interact with each other via a net electric field within the interfacial layer of a colloid. To zeroth order in the gradients, the interfacial excess number density of solute $k$ is then given by
	\begin{equation}
	n_k^\phi(r)=n_k^b\left[\exp\left(-\frac{\phi_k(r)}{k_BT}\right)-1\right],\label{eq:-61}
	\end{equation}
	where $\phi_k(r)$ is the corresponding electrostatic interaction potential.
	
	Within the Poisson-Boltzmann approximation, the chemical potential of the solute just comprises an ideal part $\mu_{k,\text{id}}(\mathbf r)$ and a contribution $\mu_{k,s}(\mathbf r)$ due to its solvation, such that $\mu_k(\mathbf r)=\mu_{k,\text{id}}(\mathbf r)+\mu_{k,s}(\mathbf r)$. In particular, the contribution $\mu_{k,s}(\mathbf r)$ is considered independent of the solute densities. As shown in appendix \ref{sec:-3}, the modified excess enthalpy density $h'_\phi(r)$ of the interfacial layer then takes the form
	\begin{equation}
	h'_\phi(r)=n_k(r)\phi^k(r)+n_k^\phi(r)k_BT^k,\label{eq:-10}
	\end{equation}
	where the index $k$ in the temperature only serves as a summation index.
	
	Within the Debye-Hückel approximation, the electrostatic interaction is assumed weak compared to the thermal energy scale, so that $\left| \phi_k(r)\right| \ll k_BT$. The potential energy $\phi_k(r)$ of the charged solute is related to the local electric potential $\varphi_E(r)$ of the interfacial layer via $\phi_k(r)=e_k\varphi_E(r)$. An expansion of eq. (\ref{eq:-61}) up to second order in $\phi_k(r)$ then allows a linearisation of the Poisson equation, which yields the well-known Yukawa form of the local electric potential
	\begin{equation}
	\varphi_E(r)=\zeta\frac{R}{r}\exp\left\{-\frac{r-R}{\lambda}\right\},\label{eq:-102}
	\end{equation}
	where $\lambda=\left[n_k^be_ke^k/(\epsilon k_BT)\right]^{-1/2}$ is the Debye screening length, $\zeta$ is the electric surface potential of the colloid (at $r=R$), and $\epsilon$ ($=\epsilon_\text{out}$) is the electric permittivity of the solvent.
	
	\subsection{Electrophoresis}
	
	Electrophoretic motion of charged colloids occurs in the presence of an applied electric field $\mathbf E_b$. Using eq. (\ref{eq:-103}) with $\nabla Y(\mathbf r)=\nabla_T\mu_k(\mathbf r)-e_k\mathbf E(\mathbf r)$, $\nabla Y_b=-e_k\mathbf E_b$ and $y'=L'_{kk}$, the local electrochemical gradients can be written as
	\begin{equation}
	\nabla_T\mu_k(\mathbf r)-e_k\mathbf E(\mathbf r)=-\mathbf C_{L'_{kk}}(\mathbf r)e_k\mathbf E_b,
	\end{equation}
	where $\mathbf C_{L'_{kk}}(\mathbf r)$ is given by eq. (\ref{eq:-97}). In view of eq. (\ref{eq:-46}), the corresponding electrophoretic force can hence be expressed as
	\begin{eqnarray}
	\mathbf F_\text{ph} &=& e^k\mathbf E_b\int_R^\infty n_k^\phi(r)B_{L'_{kk}}(r) 4\pi r^2dr\nonumber\\
	&&+L'_{kk}V_Rn_k^be^k\mathbf E_b.\label{eq:-56}
	\end{eqnarray}
	To leading order in $ \phi_k(r)$, the solute excess densities read
	\begin{equation}
	n_k^\phi(r)=-\frac{ n_k^be_k}{k_BT}\varphi_E(r).\label{eq:-11}
	\end{equation}
	With eqs. (\ref{eq:-55}), (\ref{eq:-102}) and (\ref{eq:-11}), the electrophoretic force can finally be evaluated from eq. (\ref{eq:-56}). In order to quantify the sign and strength of electrophoresis, we introduce the electrophoretic mobility $M_E$ via $\mathbf F_{\text{ph}}=\xi M_E\mathbf E_b$. Using eq. (\ref{eq:-57}) for the friction coefficient, the electrophoretic mobility takes the form
	\begin{eqnarray}
	M_E&=&\frac{2\epsilon\zeta}{3a\eta}\left(1+(1-a)x\right)\nonumber\\
	&&+L_{kk}'\frac{2\epsilon\zeta}{3a\eta}\frac{n_k^be_ke^k\lambda^2}{\epsilon k_BT}A_E(x)\nonumber\\
	&&+\frac{L'_{kk}}{6\pi a\eta R}V_Rn_k^be^k,\label{eq:-58}
	\end{eqnarray}
	where $x=R/\lambda$. The function $A_E(x)$ is given by
	\begin{eqnarray}
	A_E(x)&=&-\frac{1}{8}(2-a)x^2+\frac{1}{24}(2+3a)x^3\nonumber\\
	&&-\frac{1}{48}(2-3a)\left(x^4-x^5\right)\label{eq:-95}\\
	&&-\frac{1}{48}x^4\left(12a+(2-3a)x^2\right)e^xE_1(x),\nonumber
	\end{eqnarray}
	with
	\begin{equation}
	E_1(x)=\int_x^\infty\frac{e^{-t}}{t}dt.
	\end{equation}
	Note that $\zeta$ also depends on $x$. If all solutes have the same diffusive permeability within a given region ($L_{kk}=n_k^bT/\xi_k=L$), then they share a common Clausius-Mossotti factor $L'$. In this case, eq. (\ref{eq:-58}) reduces to
	\begin{equation}
	M_E=\frac{2\epsilon\zeta}{3a\eta}\left(1+(1-a)x+L'A_E(x)\right),\label{eq:-96}
	\end{equation}
	where we note that the contribution from the electrochemical perturbation, given by the last term in eq. (\ref{eq:-58}), has dropped out because $n_k^be^k=0$. In addition, let us consider a charged colloid with a non-slip hydrodynamic boundary ($a=1$), which is impermeable to the solutes ($L'=-1/2$). Using these values and substituting the expression for $A_E(x)$ into eq. (\ref{eq:-96}), we obtain
	\begin{equation}
	\begin{split}
	M_E=\frac{2\epsilon\zeta}{3\eta}&\left[1+\frac{1}{16}x^2-\frac{5}{48}x^3-\frac{1}{96}\left(x^4-x^5\right)\right.\\
	&\left.+\frac{1}{96}x^4\left(12-x^2\right)e^xE_1(x)\right].\label{eq:-3}
	\end{split}
	\end{equation}
	The result in brackets exactly coincides with the Henry function for the electrophoretic mobility,\cite{Henry1931,Swan2012} which was first derived by Henry using force-free rather than reciprocal arguments. As a result, the Henry function is restricted to a non-slip boundary condition and assumes that the solutes cannot diffusive through the hydrodynamic boundary of a colloid.
	
	\subsection{Thermophoresis}
	
	It is also worth enquiring how electrostatic interactions affect colloidal motion inside an applied temperature gradient $\nabla T_b$. With $\phi_k(r)=e_k\varphi_E(r)$, the interfacial excess enthalpy density becomes
	\begin{equation}
	h'_\phi(r)=\rho_e^\phi(r)\varphi_E(r)+n_k^\phi(r)k_BT^k.\label{eq:-60}
	\end{equation}
	Moreover, we have $\rho_e^\phi(r)=-\epsilon\varphi_E(r)/\lambda^2$ within the Debye-Hückel approximation, meaning that the electrostatic energy density in eq. (\ref{eq:-60}) scales with the square of the local electric potential. For the second term in eq. (\ref{eq:-60}), the expression of $n_k^\phi(r)$ given by eq. (\ref{eq:-61}) must therefore be expanded up to second order in $\phi_k(r)$. This yields
	\begin{equation}
	h'_\phi(r)=-\frac{\epsilon}{2\lambda^2}\varphi_E^2(r)\label{eq:-76}
	\end{equation}
	to leading order in $\phi_k(r)$. Given that $\nabla T(\mathbf r)=\mathbf C_{\kappa'}(\mathbf r)\cdot\nabla T_b$, eq. (\ref{eq:-46}) can then be used to determine the thermophoretic force from
	\begin{eqnarray}
	\mathbf F_{\text{ph}} = -\frac{\nabla T_b}{T}\int_R^\infty h'_\phi(r) B_{\kappa'}(r) 4\pi r^2dr-\kappa'V_Rh'_b\frac{\nabla T_b}{T}\nonumber,
	\end{eqnarray}
	where $\kappa'=(\kappa_{\text{in}}-\kappa_{\text{out}})/(\kappa_{\text{in}}+2\kappa_{\text{out}})$. However, the thermophoretic force is not the only force that drives colloidal motion inside a temperature gradient. From eqs. (\ref{eq:-36}) and (\ref{eq:-59}), we see that the osmotic pressure gradient exerts an additional force $-k_B\nabla T_b$ on the colloid. Noting that $h'_b=-ck_BT$, the net driving force $\mathbf F_{T}=\mathbf F_{\text{ph}}-k_B\nabla T_b$ induced by the temperature gradient becomes
	\begin{eqnarray}
	\mathbf F_{T}&=&-\frac{\nabla T_b}{T}\int_R^\infty h'_\phi(r) B_{\kappa'}(r) 4\pi r^2dr\nonumber\\
	&&-k_B\nabla T_b\left(1-\kappa'cV_R\right).\label{eq:-78}
	\end{eqnarray}
	Let us further introduce the 'thermal' mobility $D_T$ via $\mathbf F_T=-\xi D_T\nabla T_b$. Using eq. (\ref{eq:-76}) to evaluate the integral in eq. (\ref{eq:-78}), we obtain
	\begin{equation}
	D_T=\frac{1}{6}\frac{\epsilon\zeta^2}{a\eta T}A_{\kappa'}(x)+\frac{k_B}{6\pi a\eta R}(1-\kappa'cV_R),\label{eq:-98}
	\end{equation} 
	where $x=R/\lambda$, and
	\begin{equation}
	\begin{split}
	&A_{\kappa'}(x)=\\
	&\hspace{0.5cm}x-\frac{2}{15}\kappa'(3-2a)x^2\\
	&+\frac{1}{30}\kappa'(6+a)x^3-\frac{1}{15}\kappa'(2+7a)x^4\\
	&+\frac{1}{15}\kappa'(2-3a)\left(x^5-2x^6\right)\\
	&-\frac{1}{15}\left(30ax^2-20\kappa'ax^5-4\kappa'(2-3a)x^7\right)e^{2x}E_1(2x).
	\end{split}
	\end{equation}
	If the colloid and the solvent have the same thermal conductivity ($\kappa'=0$), the temperature gradient is uniform. A non-slip boundary condition ($a=1$) then yields
	\begin{equation}
	D_T=\frac{1}{6}\frac{\epsilon\zeta^2}{\eta T}\left[x-2x^2e^{2x}E_1(2x)\right]+\frac{k_B}{6\pi \eta R},\label{eq:-64}
	\end{equation}
	where the first term is the 'phoretic', and the second term is the 'osmotic' contribution to the thermal mobility. For dilute colloids, the osmotic contribution is expected to be negligible compared to the phoretic contribution.\cite{Piazza2008} When $x$ becomes very large, the expression in brackets in eq. (\ref{eq:-64}) tends to the value $1/2$. In the boundary-layer limit ($R\gg\lambda$), the mobility $D_T$ therefore reduces to
	\begin{equation}
	D_T=\frac{1}{12}\frac{\epsilon\zeta^2}{\eta T}+\frac{k_B}{6\pi \eta R}.\label{eq:-63}
	\end{equation}
	The first term in eq. (\ref{eq:-63}) corresponds to the well-known Ruckenstein term of the thermophoretic mobility, which was first derived from a force-free argument.\cite{Ruckenstein1981} In fact, Ruckenstein obtained a pre-factor $1/8$ instead of $1/12$, by considering a planar rather than spherical surface.
	
	The recovery of the Henry function for electrophoresis and the Ruckenstein term for thermophoresis validates eq. (\ref{eq:-46}) and suggests that our hydrodynamic form of the phoretic force is a generalisation of these results to non-uniform thermodynamic gradients ($-\frac{1}{2}\leqslant y'\leqslant 1$), for any hydrodynamic boundary condition ($\frac{2}{3}\leqslant a\leqslant 1$) or interfacial layer thickness ($\lambda\sim R$). The corresponding colloidal flux (or velocity) can hence be determined from eqs. (\ref{eq:-58}) or (\ref{eq:-98}) and compared to experimental measurements, provided that the applied thermodynamic gradients are known. In the light of these findings, it should however be remembered that the application of one thermodynamic gradient may well lead to the induction of other gradients inside the system \cite{Prieve1984,Putnam2005} (see appendix \ref{sec:-1} for further details). For instance, a favourable comparison to experimental measurements has previously been achieved by noting that a stationary temperature gradient may induce a thermoelectric field in the bulk of an aqueous electrolyte solution.\cite{Rasuli2008,Wurger2008,Burelbach2019} This suggests that colloidal motion inside multi-component fluids often results from an interplay between multiple phoretic phenomena, even if only one thermodynamic gradient is externally applied to the system.
	\\
	\section{Conclusion}
	Based on Onsager's reciprocal relations, we have formulated a complete description of colloidal motion driven by non-uniform thermodynamic forces. Our formulation shows that the colloidal flux results from a local coupling between the Stokes flows and thermodynamic forces inside the fluid, which reflects both the hydrodynamic and thermodynamic character of particle motion. The separation of the fluid into an interfacial region and a reservoir has further allowed us to identify the phoretic force density and the osmotic pressure gradient as primary driving forces behind non-equilibrium motion of colloids. Our general results are expected to apply to both passive and active colloids, as long as the suspension remains at local thermodynamic equilibrium. Moreover, our results also suggest that phoretic motion can occur in the absence of interfacial interactions, as evidenced by an example of diffusiophoresis at uniform temperature. An explicit hydrodynamic form of the phoretic force has further been derived for a sphere in the limit of high dilution. The obtained expression extends the Henry function for electrophoresis and the Ruckenstein term for thermophoresis to arbitrary hydrodynamic boundary conditions and interfacial layer thicknesses. In general, our formulation provides a common ground for all colloidal transport phenomena and closes the gap between the force-free and reciprocal approaches to non-equilibrium motion.
	
	\section{Acknowledgements}
	The author acknowledges helpful discussions with Daan Frenkel.
	
	\section{Appendix}
	
	\subsection{Derivation of the phoretic force for passive colloids\label{sec:-2}}
	
	Here, we derive the hydrodynamic form of the phoretic force for passive colloids in the dilute limit, under the assumption that the temperature and electrochemical fields satisfy the Laplace equation
	\begin{equation}
	\nabla\cdot y(\mathbf r)\nabla Y(\mathbf r)=0, \ \ y(\mathbf r)=
	\begin{cases}
	y_\text{out} & \text{if}\ r\geqslant R \\
	y_\text{in} & \text{if}\ r<R,
	\end{cases}
	\end{equation}
	where
	\begin{equation}
	y(\mathbf r)\equiv\left\{\kappa(\mathbf r),\frac{L_{kk}(\mathbf r)}{T}\right\}.
	\end{equation} 
	Requiring continuity of $y(\mathbf r)\nabla Y(\mathbf r)$ at the hydrodynamic boundary and noting that $\nabla Y(\mathbf r)\rightarrow \nabla Y_b$ far away from the colloid, the Laplace equation can be solved for a sphere of radius $R$ and transport coefficient $y_{\text{in}}$ embedded in a fluid medium with a transport coefficient $y_{\text{out}}$. The corresponding solution is well-known and given by \cite{Epstein1929}
	\begin{equation}
	Y(r,\theta)=
	\begin{cases}
	Y^*+\left|\nabla Y_b\right|\left(1-y'\frac{R^3}{r^3}\right)r\cos\theta, &r\geqslant R \\
	\\
	Y^*+\left|\nabla Y_b\right|\left(1-y'\right)r\cos\theta, &r<R,
	\end{cases}\label{eq:-77}
	\end{equation}
	where $\nabla Y_b$ is the unperturbed gradient inside the bulk fluid, $\theta$ is the angle between $\nabla Y_b$ and the position $\mathbf r$ from the colloidal centre, and $Y^*$ is a constant reference value at $\theta=\pi/2$. The Clausius-Mossotti factor reads
	\begin{equation}
	y'=\frac{y_\text{in}-y_\text{out}}{y_\text{in}+2y_\text{out}}.
	\end{equation} 
	From this solution, the local gradient $\nabla Y(\mathbf r)$ is found to be related to the applied bulk gradient $\nabla Y_b$ via
	\begin{equation}
	\nabla Y(\mathbf r)=\mathbf C_{y'}(\mathbf r)\cdot\nabla Y_b,\label{eq:-88}
	\end{equation}
	where the tensor $\mathbf C_{y'}(\mathbf r)$ is identified as
	\begin{equation}
	\mathbf C_{y'}(\mathbf r)=
	\begin{cases}
	\mathbf 1+y'\frac{R^3}{r^3}\left(3\hat {\mathbf r}\hat {\mathbf r}-\mathbf 1\right), &r\geqslant R \\
	\\
	\left(1-y'\right)\mathbf 1, &r<R.
	\end{cases}\label{eq:-43}
	\end{equation}
	
	At high dilution, a large proportion of fluid is located in the bulk region of the system, which implies that $\left\langle\nabla Y(\mathbf r)\right\rangle_{V_f}=\nabla Y_b$. Therefore, the local perturbation of $\nabla Y(\mathbf r)$ inside the fluid can be expressed as
	\begin{equation}
	\delta\nabla Y(\mathbf r)=\Theta(\mathbf r)\left(\mathbf C_{y'}(\mathbf r)-\mathbf 1\right)\cdot\nabla Y_b,\label{eq:-75}
	\end{equation}
	where $\Theta(\mathbf r)$ represents the excluded volume function for the colloidal volume $V_c$. Noting that $\mathbf S(\mathbf r)-\mathbf 1=\mathbf 0$ for $r<R$, eq. (\ref{eq:-26}) can now be written as
	\begin{equation}
	\mathbf F_{Y} = -\nabla Y_b\cdot\int_R^\infty\left[\rho_\phi(\mathbf r)\mathbf B(\mathbf r)+\rho_b\mathbf B^*(\mathbf r)\right]dV,\label{eq:-48}
	\end{equation}
	where the tensors $\mathbf B(\mathbf r)$ and $\mathbf B^*(\mathbf r)$ are given by
	\begin{equation}
	\mathbf B(\mathbf r)=\left(\mathbf S(\mathbf r)-\mathbf 1\right)\cdot\mathbf C_{y'}(\mathbf r)
	\end{equation}
	and
	\begin{equation}
	\mathbf B^*(\mathbf r)=\left(\mathbf S(\mathbf r)-\mathbf 1\right)\cdot\left(\mathbf C_{y'}(\mathbf r)-\mathbf 1\right).
	\end{equation}
	The notation $\int_R^\infty$ indicates that the volume integral is evaluated from the hydrodynamic boundary at $r=R$ to a bulk region far away from the boundary ($r\rightarrow\infty$).
	
	For our model calculations, we assume that the interfacial excess density $\rho_\phi(r)$ only depends on radial distance. In this case, only the $z$-component of the integrand contributes to the volume integral, due to the rotational symmetry around the direction $\hat {\mathbf z}$ of $\nabla Y_b$, hence
	\begin{equation}
	\mathbf F_{Y}=-\nabla Y_b\int_R^\infty\left[\rho_\phi(r) B_z(r,\theta)+\rho_bB_z^*(r,\theta)\right]dV,
	\end{equation}
	where the scalar functions $B_z(r,\theta)=\mathbf B(\mathbf r):\hat {\mathbf z}\hat {\mathbf z}$ and $B_z^*(r,\theta)=\mathbf B^*(\mathbf r):\hat {\mathbf z}\hat {\mathbf z}$ are the projections of these tensors onto $\hat {\mathbf z}$. Based on eqs. (\ref{eq:-42}) and (\ref{eq:-43}), the expression of $B_z(r,\theta)$ is given by
	\begin{equation}
	\begin{split}
	B_z(r,\theta)=&-\cos^2\theta\left[1+2y'\frac{R^3}{r^3}\right]\\
	&\hspace{0.9cm}\times\left[1-\frac{3a}{2}\frac{R}{r}-\left(1-\frac{3a}{2}\right)\frac{R^3}{r^3}\right]\\
	&-\sin^2\theta\left[1-y'\frac{R^3}{r^3}\right]\\
	&\hspace{0.9cm}\times\left[1-\frac{3a}{4}\frac{R}{r}+\frac{1}{2}\left(1-\frac{3a}{2}\right)\frac{R^3}{r^3}\right].
	\end{split}
	\end{equation}
	By introducing the orientational average
	$\left\langle B_z(r)\right\rangle_\theta=\frac{1}{2}\int_{0}^{\pi}B_z(r,\theta)\sin\theta d\theta$, we can then write
	\begin{eqnarray}
	\mathbf F_{Y}&=&-\nabla Y_b\int_R^\infty\rho_\phi(r) \left\langle B_z(r)\right\rangle_\theta 4\pi r^2dr\nonumber\\
	&&-\rho_b\nabla Y_b\int_R^\infty\left\langle B_z^*(r)\right\rangle_\theta 4\pi r^2dr,\label{eq:-45}
	\end{eqnarray}
	where
	\begin{equation}
	\left\langle B_z(r)\right\rangle_\theta=-1+a\frac{R}{r}+y'\left[\frac{a}{2}\frac{R^4}{r^4}+\left(1-\frac{3a}{2}\right)\frac{R^6}{r^6}\right].\label{eq:-44}
	\end{equation}
	The expression for $\left\langle B_z^*(r)\right\rangle_\theta$ can directly be obtained from eq. (\ref{eq:-44}) by only keeping the terms proportional to $y'$: 
	\begin{equation}
	\left\langle B_z^*(r)\right\rangle_\theta=y'\left[\frac{a}{2}\frac{R^4}{r^4}+\left(1-\frac{3a}{2}\right)\frac{R^6}{r^6}\right].
	\end{equation}
	The second integral in eq. (\ref{eq:-45}) can now be evaluated, giving $-y'V_R\rho_b\nabla Y_b$, which is independent of the hydrodynamic boundary condition. Hence, the contribution $\mathbf F_{Y}$ to the phoretic force takes the final form
	\begin{equation}
	\mathbf F_{Y}=-\nabla Y_b\int_R^\infty\rho_\phi(r) \left\langle B_z(r)\right\rangle_\theta 4\pi r^2dr-y'V_R\rho_b\nabla Y_b,\label{eq:-47}
	\end{equation}
	which, by denoting $\left\langle B_z(r)\right\rangle_\theta\equiv B_{y'}(r)$, corresponds to our result given by eq. (\ref{eq:-46}).
	
	Equation (\ref{eq:-46}) describes the phoretic force for passive colloids. For self-phoretic motion of active colloids, no thermodynamic gradients are externally applied to the system. The gradient $\nabla Y(\mathbf r)$ must therefore stem from a source density $\sigma(\mathbf r)$ generated by the colloid, and the corresponding field satisfies the Poisson equation $\nabla\cdot y(\mathbf r)\nabla Y(\mathbf r)=-\sigma(\mathbf r)$, where $\sigma(\mathbf r)\equiv\left\{\sigma_q(\mathbf r),\sigma_k(\mathbf r)\right\}$. This equation does usually not have a straightforward analytical solution when transport properties differ inside and outside the boundary.\cite{Bickel2013} However, a particularly simple case is given by that of a self-phoretic dipole, with a corresponding dipole moment $\mathbf p$. From eq. (\ref{eq:-43}), we see that the outer perturbation caused by a sphere inside a uniform bulk gradient is that of a dipole with moment $\mathbf p=-3yy'V_R\nabla Y_b$. If the interfacial layer is again assumed to be spherically symmetric, the self-phoretic force of a dipole can hence directly be obtained from eqs. (\ref{eq:-46}) and (\ref{eq:-55}) by setting $y'\nabla Y_b=-\mathbf p/(3yV_R)$
	and omitting all other terms, giving
	\begin{equation}
	\begin{split}
	\mathbf F_{Y}&=\\
	&\frac{\mathbf p}{y R}\int_R^\infty\rho_\phi(r) \left[\frac{a}{2}\frac{R^2}{r^2}+\left(1-\frac{3a}{2}\right)\frac{R^4}{r^4}\right]dr\nonumber+\frac{\rho_b\mathbf p}{3y}.
	\end{split}
	\end{equation}	
	The self-phoretic force of a dipole thus takes a non-zero value $\rho_b\mathbf p/(3y)$ in the absence of interfacial excess.
	
	\subsection{Determination of the thermodynamic fields\label{sec:-1}}
	
	In order to determine the local temperature field and electrochemical fields for passive colloids in the dilute limit, we have to make use of the stationary continuity equations 
	\begin{equation}
	\nabla\cdot\mathbf J'_q(\mathbf r)=\sigma_q(\mathbf r)\label{eq:-54}
	\end{equation}
	and
	\begin{equation}
	\nabla\cdot\mathbf J_k(\mathbf r)=\sigma_k(\mathbf r),
	\end{equation}
	where $\sigma_q(\mathbf r)$ and $\sigma_k(\mathbf r)$ are the corresponding source (or sink) densities. As the local heat and fluid particle fluxes are given by eqs. (\ref{eq:-39}) and (\ref{eq:-38}), solving these equations for the thermodynamic fields is in general not trivial, even in the absence of sources. It has been suggested by Anderson\cite{Anderson1989} that the cross-coefficients of these fluxes are negligible compared to the diagonal coefficients, and that the latter may be assumed uniform inside and outside the hydrodynamic boundary of a colloid. Indeed, the hydrodynamic contributions in eqs. (\ref{eq:-39}) and (\ref{eq:-38}) can usually be neglected for the determination of the thermodynamic fields due to the relatively slow motion of the colloids. Under this assumption, the fluxes reduce to
	\begin{eqnarray}
	\mathbf{J}'_{q}(\mathbf r) & = & -L_{qq}(\mathbf r)\frac{\nabla T(\mathbf r)}{T^2}\nonumber\\
	&&+\frac{1}{T}L_{ql}(\mathbf r)\left( -\nabla_T\mu^l(\mathbf r)+\mathbf{F}^l(\mathbf r)\right)\nonumber
	\end{eqnarray}
	and
	\begin{eqnarray}
	\mathbf{J}_{k}(\mathbf r) & = & -L_{kq}(\mathbf r)\frac{\nabla T(\mathbf r)}{T^2}\nonumber\\
	&&+\frac{1}{T}L_{kl}(\mathbf r)\left( -\nabla_T\mu^l(\mathbf r)+\mathbf{F}^l(\mathbf r)\right)\nonumber.
	\end{eqnarray}
	As heat transport is dominated by conduction in colloidal suspensions ($L_{qq}\gg k_BTL_{ql}$), we have $\mathbf{J}'_{q}(\mathbf r)\approx-\kappa(\mathbf r)\nabla T(\mathbf r)$, where $\kappa(\mathbf r)=L_{qq}(\mathbf r)/T^2$ is the thermal conductivity. Eq. (\ref{eq:-54}) therefore simplifies to
	\begin{equation}
	\nabla\cdot\kappa(\mathbf r)\nabla T(\mathbf r)=-\sigma_q(\mathbf r),\label{eq:-13}
	\end{equation}
	with
	\begin{equation}
	\kappa(\mathbf r)=
	\begin{cases}
	\kappa_\text{out} & \text{if}\ r\geqslant R \\
	\kappa_\text{in} & \text{if}\ r<R.
	\end{cases}
	\end{equation}
	Due to the incompressibility and high density of the solvent ($k=0$), the solvent flux is well approximated by $\mathbf J_0(\mathbf r)\approx\frac{L_{00}(\mathbf r)}{T}\left(-\nabla_T\mu_0(\mathbf r)+\mathbf F_{g,0}\right)$. Denoting the corresponding source density by $\sigma_0(\mathbf r)$, the continuity equation for the solvent becomes 
	\begin{equation}
	\nabla\cdot\frac{L_{00}(\mathbf r)}{T}\left(\nabla_T\mu_0(\mathbf r)-\mathbf F_{g,0}\right)=-\sigma_0(\mathbf r),\label{eq:-14}
	\end{equation}
	with
	\begin{equation}
	L_{00}(\mathbf r)=
	\begin{cases}
	L_{00,\text{out}} & \text{if}\ r\geqslant R \\
	L_{00,\text{in}} & \text{if}\ r<R.
	\end{cases}
	\end{equation}
	
	In order to justify the constancy of the Onsager coefficients $L_{kk}(\mathbf r)$ and $L_{kq}(\mathbf r)$ of the remaining solutes ($k\neq 0$) outside the hydrodynamic boundary, the excess densities $n_k^\phi(\mathbf r)$ of the interfacial layer must be assumed weak compared to the bulk densities, so that $n_k(\mathbf r)\approx n_k^b$ for $r\geqslant R$. This condition is satisfied in the Debye-Hückel approximation, where the interfacial interaction potential of the solutes is assumed weak compared to the thermal energy. Moreover, cross-coupling to other solute components can be neglected within the Poisson-Boltzmann approximation ($L_{k l}\approx 0$ for $l\neq k$).\cite{Burelbach2019} However, the 'heat of transport' $Q^*_{k}=L_{k q}(\mathbf r)/L_{kk}(\mathbf r)$ of the solute may  be comparable or larger than the thermal energy $k_BT$,\cite{Wurger2010} meaning that the cross-coefficient $L_{kq}(\mathbf r)$ cannot be neglected for the solute flux.  Noting that $Q^*_{k}$ is a constant single-particle (solvation) property within the Poisson-Boltzmann approximation,\cite{Agar1989,Burelbach2019} and ignoring the weight of the solutes, the solute flux takes the form
	\begin{equation}
	\mathbf J_k(\mathbf r)\approx\frac{L_{kk}(\mathbf r)}{T}\left(-Q^*_{k}\frac{\nabla T(\mathbf r)}{T}-\nabla_T\mu_k(\mathbf r)+e_k\mathbf E(\mathbf r)\right).\label{eq:-90}
	\end{equation} 
	Accordingly, the continuity equation for solute component $k$ is given by
	\begin{eqnarray}
	\nabla\cdot\frac{L_{kk}(\mathbf r)}{T}\left(Q^*_{k}\frac{\nabla T(\mathbf r)}{T}+\nabla_T\mu_k(\mathbf r)-e_k\mathbf E(\mathbf r)\right)\nonumber\\
	=-\sigma_k(\mathbf r),\nonumber\\
	\label{eq:-19}
	\end{eqnarray}
	where
	\begin{equation}
	L_{kk}(\mathbf r)=
	\begin{cases}
	L_{kk,\text{out}} & \text{if}\ r\geqslant R \\
	L_{kk,\text{in}} & \text{if}\ r<R.
	\end{cases}
	\end{equation}
	
	For passive colloids, we have $\sigma_q(\mathbf r)=0$ and $\sigma_k(\mathbf r)=0$, and eqs. (\ref{eq:-13}), (\ref{eq:-14}) and (\ref{eq:-19}) reduce to the following set of Laplace equations
	\begin{equation}
	\nabla\cdot\kappa(\mathbf r)\nabla T(\mathbf r)=0,\label{eq:-66}
	\end{equation}
	\begin{equation}
	\nabla\cdot \frac{L_{00}(\mathbf r)}{T}\left(\nabla_T\mu_0(\mathbf r)-\mathbf F_{g,0}\right)=0\label{eq:-69}
	\end{equation}
	and
	\begin{equation}
	\nabla\cdot \frac{L_{kk}(\mathbf r)}{T}\left(Q^*_{k}\frac{\nabla T(\mathbf r)}{T}+\nabla_T\mu_k(\mathbf r)-e_k\mathbf E(\mathbf r)\right)=0,\label{eq:-74}
	\end{equation}
	where the last equation applies to the solutes ($k\neq0$). As a result, the solutions of these equations are given by eq. (\ref{eq:-77}). From eqs. (\ref{eq:-66})-(\ref{eq:-74}), we also see that only the temperature field directly satisfies the Laplace equation, whereas the electrochemical fields of the solutes additionally couple to the temperature field. Based on eq. (\ref{eq:-88}), the thermodynamic gradients defined in eq. (\ref{eq:-50}) can now be expressed in terms of those occurring in eqs. (\ref{eq:-66})-(\ref{eq:-74}), yielding the relations
	\begin{equation}
	\nabla T(\mathbf r)=\mathbf C_{\kappa'}(\mathbf r)\cdot\nabla T_b,\label{eq:-93}
	\end{equation}
	\begin{equation}
	\nabla_T\mu_0(\mathbf r)=\mathbf C_{L_{00}'}(\mathbf r)\cdot\nabla_T\mu_0^b-\left(\mathbf C_{L_{00}'}(\mathbf r)-\mathbf 1\right)\cdot\mathbf F_{g,0},\label{eq:-94}
	\end{equation}
	and
	\begin{eqnarray}
	\nabla_T\mu_k(\mathbf r)-e_k\mathbf E(\mathbf r)&& \ =\label{eq:-89}\\
	&&\mathbf C_{L_{kk}'}(\mathbf r)\cdot\left(\nabla_T\mu_k^b-e_k\mathbf E_b\right)\nonumber\\
	&&+Q_{k}^*\left(\mathbf C_{L_{kk}'}(\mathbf r)-\mathbf C_{\kappa'}(\mathbf r)\right)\cdot\frac{\nabla T_b}{T}.\nonumber
	\end{eqnarray}
	For convenience, we have used $y'=L_{kk}'$ instead of $y'=(L_{kk}/T)'$ in eqs. (\ref{eq:-94}) and (\ref{eq:-89}), as a factor $1/T$ simply drops out of the corresponding Clausius-Mossotti factor in eq. (\ref{eq:-43}). 
	
	The incompressible solvent can only contribute to the phoretic force via a perturbation of its chemical potential gradient. Using eq. (\ref{eq:-94}), this perturbation can be expressed as
	\begin{equation}
	\delta\nabla_T\mu_0(\mathbf r)=\Theta(\mathbf r)\left(\mathbf C_{L_{00}'}(\mathbf r)-\mathbf 1\right)\cdot\left(\nabla_T\mu_0^b-\mathbf F_{g,0}\right).\label{eq:-85}
	\end{equation}
	Instead of $\nabla Y_b=\nabla_T\mu_0^b$, the substitution $\nabla Y_b=\nabla_T\mu_0^b-\mathbf F_{g,0}$ must therefore be used for the solvent under gravity, when evaluating the second term in eq. (\ref{eq:-46}).
	
	In view of eq. (\ref{eq:-89}), there are different scenarios for the phoretic motion of colloids due to electrochemical forces on the solutes, depending on which gradients are applied, and how these gradients are applied to the system.\cite{Henry1931,Brady2011,Prieve1984,Burelbach2019} 
	The scenario considered in section \ref{sec:-7} holds for systems where electrochemical forces only occur in the absence of a temperature gradient, or $vice$ $versa$. In this case, eq. (\ref{eq:-89}) reduces to
	\begin{equation}
	\nabla_T\mu_k(\mathbf r)-e_k\mathbf E(\mathbf r)=\mathbf C_{L_{kk}'}(\mathbf r)\cdot\left(\nabla_T\mu_k^b-e_k\mathbf E_b\right),
	\end{equation}
	meaning that the electrochemical fields also directly satisfy the Laplace equation, as required by eq. (\ref{eq:-40}).
	This scenario therefore corresponds to diffusio- and electrophoretic motion at uniform temperature,\cite{Henry1931,Brady2011,Prieve1984} or conversely to thermophoresis in the absence of electrochemical forces.\cite{Parola2004} For the case of electrophoresis presented in section \ref{sec:-7}, we have also assumed that the application of an electric field does not induce any solute chemical potential gradients in the bulk of the system.\cite{Henry1931} This condition is satisfied if the electric field is periodically reversed, as to avoid the accumulation of solutes on one or the other side of the system. A different situation arises when the bulk is subjected to stationary chemical gradients of charged solutes, which may induce an electric field $\mathbf E_b$ by solute diffusion.\cite{Prieve1984} In this case, $\mathbf E_b$ can be determined from the condition of vanishing electric current in the bulk of the system ($e^k\mathbf J_k^b=0$).
	
	Another rather common scenario, which is not considered in detail here, occurs when the solutes reach a steady state inside an applied temperature gradient, such that $\mathbf J_k^b=0$.\cite{Wurger2008} Based on eq. (\ref{eq:-90}), this condition yields
	\begin{equation}
	\nabla_T\mu_k^b-e_k\mathbf E_b=-Q^*_{k}\frac{\nabla T_b}{T}.\label{eq:-91}
	\end{equation}
	Equation (\ref{eq:-91}) can then be used to eliminate the induced electrochemical bulk gradients in eq. (\ref{eq:-89}), giving
	\begin{eqnarray}
	\nabla_T\mu_k(\mathbf r)-e_k\mathbf E(\mathbf r)=-Q_{k}^*\mathbf C_{\kappa'}(\mathbf r)\cdot\frac{\nabla T_b}{T}.\label{eq:-92}
	\end{eqnarray}
	If the colloids have the same thermal conductivity as the solvent, then $\mathbf C_{\kappa'}(\mathbf r)=\mathbf 1$ and the temperature gradient is uniform everywhere. In this case, eq. (\ref{eq:-92}) reduces to $\nabla_T\mu_k-e_k\mathbf E=-Q_{k}^*\nabla T_b/T$, which has previously been used to describe the effect of electrochemical gradients on thermophoresis inside a stationary temperature gradient.\cite{Burelbach2019,Wurger2008,Putnam2005}
	
	\subsection{The partial enthalpy and modified enthalpy density of a point-like solute component\label{sec:-3}}
	
	Within the Poisson-Boltzmann approximation, we treat the solutes as point-like particles that do not interact with each other inside the neutral bulk of the system. The solutes are embedded in a solvent and may be surrounded by solvation shells due to their specific interaction with the solvent molecules. In this case, the chemical potential of a solute component $k$ just comprises an ideal part $\mu_{k,\text{id}}(\mathbf r)$ and a contribution $\mu_{k,s}(\mathbf r)$ due to its solvation, such that $\mu_k(\mathbf r)=\mu_{k,\text{id}}(\mathbf r)+\mu_{k,s}(\mathbf r)$. The ideal part $\mu_{k,\text{id}}(\mathbf r)$ is given by
	\begin{equation}
	\mu_{k,\text{id}}(\mathbf r)=k_BT(\mathbf r)\left(\ln n_k^r(\mathbf r) - \frac{3}{2}\ln T(\mathbf r)+K\right),\label{eq:sim-12}
	\end{equation}
	where $n_k^r(\mathbf r)$ is the corresponding reservoir density and $K$ is a solute-specific constant. As the solvent is incompressible and as $\mu_{k,s}(\mathbf r)$ is independent of the solute densities, the Gibbs adsorption equation for the solvation shell reads \cite{Burelbachthesis}
	\begin{equation}
	-d\mu_{k,s}(\mathbf r)=S_{k,s}(\mathbf r)dT(\mathbf r),
	\end{equation}
	where $S_{k,s}(\mathbf r)$ is the entropy of solvation. The partial molar enthalpy of solute component $k$ can now be computed, yielding
	\begin{eqnarray}
	\bar{H}_{k}(\mathbf r) & = & -T^2(\mathbf r)\frac{\partial}{\partial T}\left(\frac{\mu_{k}(\mathbf r)}{T(\mathbf r)} \right)_{P,n_j}\nonumber\\
	& = & -T^2(\mathbf r)\frac{\partial}{\partial T}\left(\frac{\mu_{k,\text{id}}(\mathbf r)}{T(\mathbf r)}+\frac{\mu_{k,s}(\mathbf r)}{T(\mathbf r)}\right)_{P,n_j}\nonumber\\
	& = & \frac{3}{2}k_BT(\mathbf r)+\left(T(\mathbf r)S_{k,s}(\mathbf r)+\mu_{k,s}(\mathbf r)\right)\nonumber\\
	& = & \frac{3}{2}k_BT(\mathbf r)+\bar{H}_{k,s}(\mathbf r),
	\end{eqnarray}
	where $\bar{H}_{k,s}(\mathbf r)=T(\mathbf r)S_{k,s}(\mathbf r)+\mu_{k,s}(\mathbf r)$ is the contribution from the solvation shell.
	Clearly, a position-dependence in $\bar{H}_{k}(\mathbf r)$ can only stem from a temperature variation. However, the enthalpy density is evaluated to zeroth order in the gradients, meaning that we can simply write $\bar{H}_{k}=\frac{3}{2}k_BT+\bar{H}_{k,s}$. 
	
	Let $\phi_k(r)$ be the interfacial interaction potential of solute $k$ at the colloidal surface. The corresponding local enthalpy density can be expressed as \cite{Burelbachthesis}
	\begin{eqnarray}
	h_k(r) & = &n_{k}(r)\left(\phi_k(r)+\frac{5}{2}k_BT+\bar{H}_{k,s}\right),\nonumber
	\end{eqnarray}
	where $\frac{5}{2}k_BT$ is the ideal-gas contribution.
	Given that the interaction potential vanishes far away from the colloidal surface, the reservoir enthalpy density of the solute reads
	\begin{eqnarray}
	h_k^b & = &n_{k}^b\left( \frac{5}{2}k_BT+\bar{H}_{k,s}\right)
	\end{eqnarray}
	to zeroth order in the gradients. Hence, the interfacial enthalpy density $h'_{\phi,k}(r)$ of solute component $k$ takes the form
	\begin{eqnarray}
	h_{\phi,k}'(r) & = & h'_{k}(r)-h_{k}'^b\nonumber\\
	& = & \left( h_k(r)-n_{k}(r)\bar{H}_k\right)-\left( h_k^b-n_{k}^b\bar{H}_k\right)\nonumber\\
	& = & n_{k}(r)\phi_{k}(r)+n_{k}^{\phi}(r)k_{B}T.\label{eq:th-29}
	\end{eqnarray}
	
	The net interfacial enthalpy density, as given by eq. (\ref{eq:-10}), is then obtained by summing eq. (\ref{eq:th-29}) over all solute components.
	
	\bibliography{Ref}

\end{document}